# Bistability between π-diradical open-shell and closed-shell states in indeno[1,2-*a*]fluorene

Shantanu Mishra[1]*, Manuel Vilas-Varela[2], Leonard-Alexander Lieske[1], Ricardo Ortiz[3], Shadi Fatayer[4], Igor Rončević[5], Florian Albrecht[1], Thomas Frederiksen[3,6], Diego Peña[2]* and Leo Gross[1]*

[1]IBM Research Europe – Zurich, 8803 Rüschlikon, Switzerland

[2]Center for Research in Biological Chemistry and Molecular Materials (CiQUS) and Department of Organic Chemistry, University of Santiago de Compostela, 15782 Santiago de Compostela, Spain

[3]Donostia International Physics Center (DIPC), 20018 Donostia-San Sebastián, Spain

[4]Physical Science and Engineering Division, King Abdullah University of Science and Technology (KAUST), 23955-6900 Thuwal, Saudi Arabia

[5]Department of Chemistry, University of Oxford, Oxford OX1 3TA, United Kingdom

[6]Ikerbasque, Basque Foundation for Science, 48013 Bilbao, Spain

*E-mail: SHM@zurich.ibm.com (Shantanu Mishra), diego.pena@usc.es (Diego Peña) and LGR@zurich.ibm.com (Leo Gross).

**Indenofluorenes are non-benzenoid conjugated hydrocarbons that have received great interest owing to their unusual electronic structure and potential applications in non-linear optics and photovoltaics. Here, we report the generation of unsubstituted indeno[1,2-*a*]fluorene, the final and yet unreported parent indenofluorene isomer, on various surfaces by cleavage of two C-H bonds in 7,12-dihydroindeno[1,2-*a*]fluorene through voltage pulses applied by the tip of a combined scanning tunneling microscope and atomic force microscope. On bilayer NaCl on Au(111), indeno[1,2-*a*]fluorene is in the neutral charge state, while it exhibits charge bistability between neutral and anionic states on the lower work function surfaces of bilayer NaCl on Ag(111) and Cu(111). In the neutral state, indeno[1,2-*a*]fluorene exhibits either of two ground states: an open-shell π-diradical state, predicted to be a triplet by density functional and multireference many-body perturbation theory calculations, or a closed-shell state with a *para*-quinodimethane moiety in the *as*-indacene core. Switching between open- and closed-shell states of a single molecule is observed by changing its adsorption site on NaCl.**

The inclusion of non-benzenoid carbocyclic rings is a viable route to tune the physicochemical properties of polycyclic conjugated hydrocarbons (PCHs)[1–3]. Non-benzenoid polycycles may lead to local changes in strain, conjugation, aromaticity, and, relevant to the context of the present work, induce an open-shell ground state of the corresponding PCHs[4–7]. Many non-benzenoid PCHs are also non-alternant, where the presence of odd-membered polycycles breaks the bipartite symmetry of the molecular network[8]. Figure 1a shows classical examples of non-benzenoid non-alternant PCHs, namely, pentalene, azulene and heptalene. Whereas azulene is a stable PCH exhibiting Hückel aromaticity, pentalene and heptalene are highly reactive Hückel antiaromatic and non-aromatic compounds, respectively. Benzinterposition of pentalene generates indacenes, consisting of two isomers *s*-indacene and *as*-indacene (Fig. 1b). Apart from being antiaromatic, indacenes also contain proaromatic quinodimethane (QDM) moieties (Fig. 1c)[9], which endows them with potential open-shell character. While the parent *s*-indacene and *as*-indacene have never been isolated, kinetically and thermodynamically stabilized derivatives of *s*-indacene have been synthesized[10–12]. A feasible strategy to isolate congeners of otherwise unstable non-benzenoid non-alternant PCHs is through fusion of benzenoid rings at the ends of the π-system, that is, benzannelation. For example, while the parent pentalene is highly reactive, the benzannelated congener indeno[2,1-*a*]indene is stable under ambient conditions (Fig. 1b)[13]. However, the position of benzannelation is crucial for stability: although indeno[2,1-*a*]indene is stable, its isomer indeno[1,2-*a*]indene (Fig. 1b) oxidizes under ambient conditions[14]. Similarly, benzannelation of indacenes gives rise to the family of PCHs known as indenofluorenes (Fig. 1d), which constitute the topic of the present work. Depending on the benzannelation position and the indacene core, five isomers can be constructed, namely, indeno[2,1-*b*]fluorene (**1**), indeno[1,2-*b*]fluorene (**2**), indeno[2,1-*a*]fluorene (**3**), indeno[2,1-*c*]fluorene (**4**) and indeno[1,2-*a*]fluorene (**5**).

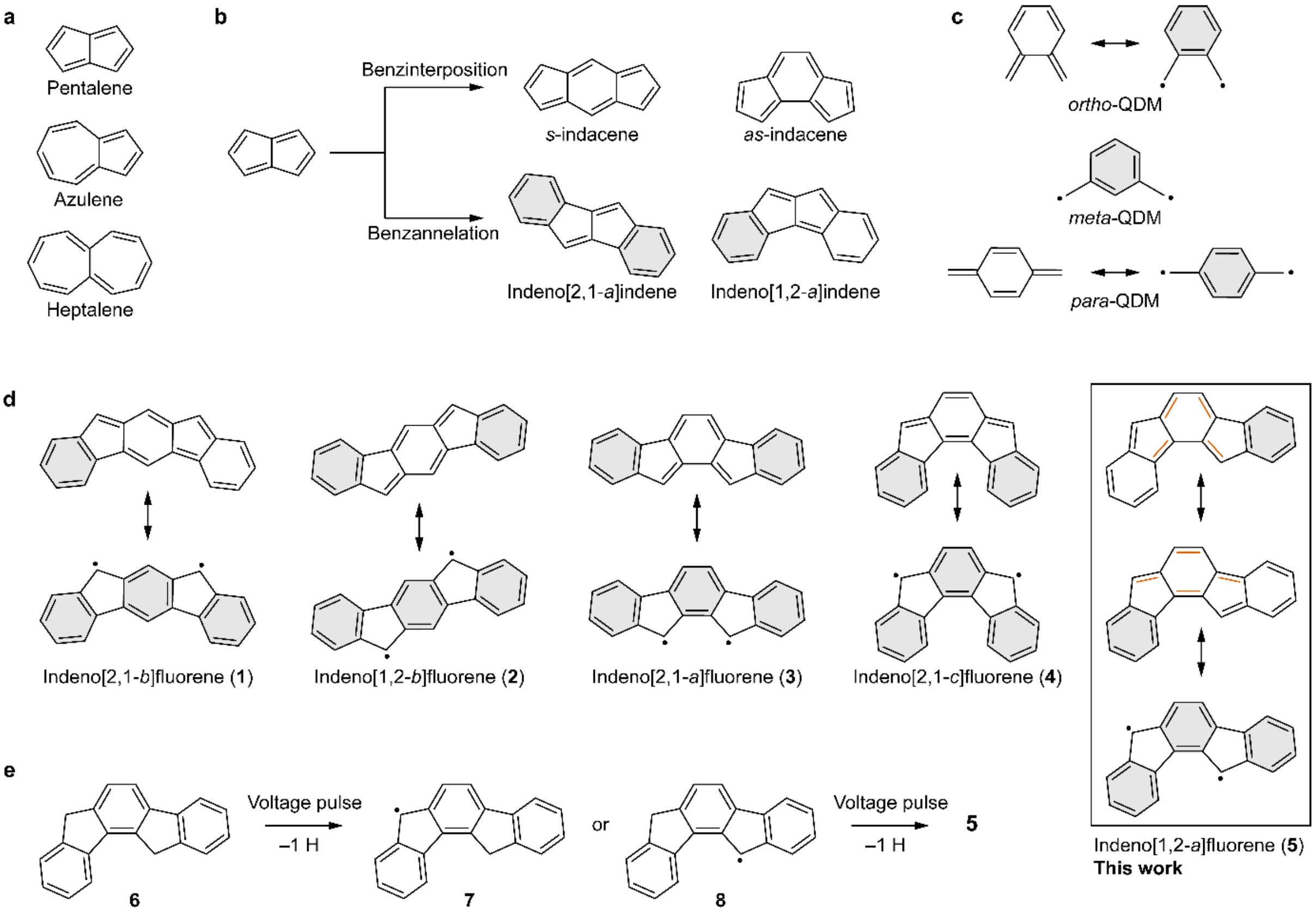

**Fig. 1 | Non-benzenoid non-alternant polycyclic conjugated hydrocarbons. a,** Classical non-benzenoid non-alternant polycyclic conjugated hydrocarbons: pentalene, azulene and heptalene. **b,** Generation of indacenes and indenoindenes through benzinterposition and benzannelation of pentalene, respectively. Gray filled rings represent Clar sextets. **c,** Closed-shell Kekulé (left) and open-shell non-Kekulé (right) resonance structures of QDMs. Note that *meta*-QDM is a non-Kekulé molecule. All indenofluorene isomers, being derived through benzannelation of indacenes, contain a central QDM moiety. **d,** Closed-shell Kekulé (top) and open-shell non-Kekulé (bottom) resonance structures of indenofluorenes. Compared to their closed-shell structures, **1** and **5** gain two Clar sextets in the open-shell structure, while **2**–**4** gain only one Clar sextet in the open-shell structure. Colored bonds in **d** highlight the *ortho*- and *para*-QDM moieties in the two closed-shell Kekulé structures of **5**. **e,** Scheme of on-surface generation of **5** ($C_{20}H_{12}$) by voltage pulse-induced dehydrogenation of **6** ($C_{20}H_{14}$). Structures **7** and **8** represent the two monoradical species ($C_{20}H_{13}$).

Practical interest in indenofluorenes stems from their low frontier orbital gap and excellent electrochemical characteristics that render them as useful components in organic electronic devices[15]. The potential open-shell character of indenofluorenes has led to several theoretical studies on their use as non-linear optical materials[16,17] and as candidates for singlet fission in organic photovoltaics[18,19]. Recent theoretical work has also shown that indenofluorene-based ladder polymers may exhibit fractionalized excitations.[20] Fundamentally, indenofluorenes represent model systems to study the interplay between aromaticity and magnetism at the molecular scale[17]. Motivated by many of these prospects, the last decade has witnessed intensive synthetic efforts toward the realization of indenofluorenes. Derivatives of **1–4** have been realized in solution[21–27], while **1–3**[28–31] have also been synthesized on surfaces and characterized using scanning tunneling microscopy (STM) and atomic force microscopy (AFM), which provide information on molecular orbital densities[32], molecular structure[33,34] and oxidation state[35,36]. With regards to the open-shell character of indenofluorenes, **2–4** are theoretically and experimentally interpreted to be closed-shell, while calculations indicate that **1** and **5** should exhibit open-shell ground states[17,28,37]. Bulk characterization of mesityl-substituted **1**, including X-ray crystallography, temperature-dependent NMR, and electron spin resonance spectroscopy, provided indications of its open-shell ground state[21]. Electronic characterization of **1** on Au(111) surface using scanning tunneling spectroscopy (STS) revealed a low electronic gap of 0.4 eV (ref. [28]). However, no experimental proof of an open-shell ground state of **1** on Au(111), such as detection of orbital densities of singly occupied molecular orbitals (SOMOs)[38,39], or spin excitations and correlations due to unpaired electrons[40,41], was shown.

In this work, we report the generation and characterization of unsubstituted **5**. Our research is motivated by theoretical calculations that indicate **5** to exhibit the largest diradical character among all indenofluorene isomers[37]. The same calculations also predict that **5** should possess a triplet ground state. Therefore, **5** would qualify as a Kekulé triplet, of which only a handful of examples exist[42–44]. However, definitive synthesis of **5** has never been reported so far. Previously, Dressler et al. reported transient isolation of mesityl-substituted **5**, where it decomposed both in the solution and in solid state[37], and only the structural proof of the corresponding dianion was obtained. On-surface generation of a derivative of **5**, starting from truxene as a precursor, was recently reported[45,46]. STM data on this compound, containing the indeno[1,2-*a*]fluorene moiety as part of a larger PCH, was interpreted to indicate its open-shell ground state[46]. However, the results did not imply the ground state of unsubstituted **5**. Here, we show that on insulating surfaces **5** can exhibit either of two ground states: an open-shell or a closed-shell. We infer the existence of these two ground states based on high-resolution AFM imaging with bond-order discrimination[34] and STM imaging of molecular orbital densities[32]. AFM imaging reveals molecules with two different geometries. Characteristic bond-order differences in the two geometries concur with the geometry of either an open- or a closed-shell state. Concurrently, STM images at ionic resonances show molecular orbital densities corresponding to SOMOs for the open-shell geometry, but orbital densities of the highest occupied molecular orbital (HOMO) and lowest unoccupied molecular orbital (LUMO) for the closed-shell geometry. Our experimental results are in good agreement with density functional theory (DFT) and multireference perturbation theory calculations. Finally, we observe switching between open- and closed-shell states of a single molecule by changing its adsorption site on the surface.

## Results and discussion

**Synthetic strategy toward indeno[1,2–*a*]fluorene.** The generation of **5** relies on the solution-phase synthesis of the precursor 7,12-dihydroindeno[1,2-*a*]fluorene (**6**). Details on synthesis and characterization of **6** are reported in Supplementary Figs. 1–3. Single molecules of **6** are deposited on coinage metal (Au(111), Ag(111) and Cu(111)) or insulator surfaces. In our work, insulating surfaces correspond to two monolayer–thick (denoted as bilayer) NaCl on coinage metal surfaces. Voltage pulses ranging between 4–6 V are applied by the tip of a combined STM/AFM system, which result in cleavage of one C-H bond at each of the pentagonal apices of **6**, thereby leading to the generation of **5** (Fig. 1e). In the main text, we focus on the generation and characterization of **5** on insulating surfaces. Generation and characterization of **5** on coinage metal surfaces is shown in Supplementary Fig. 4.

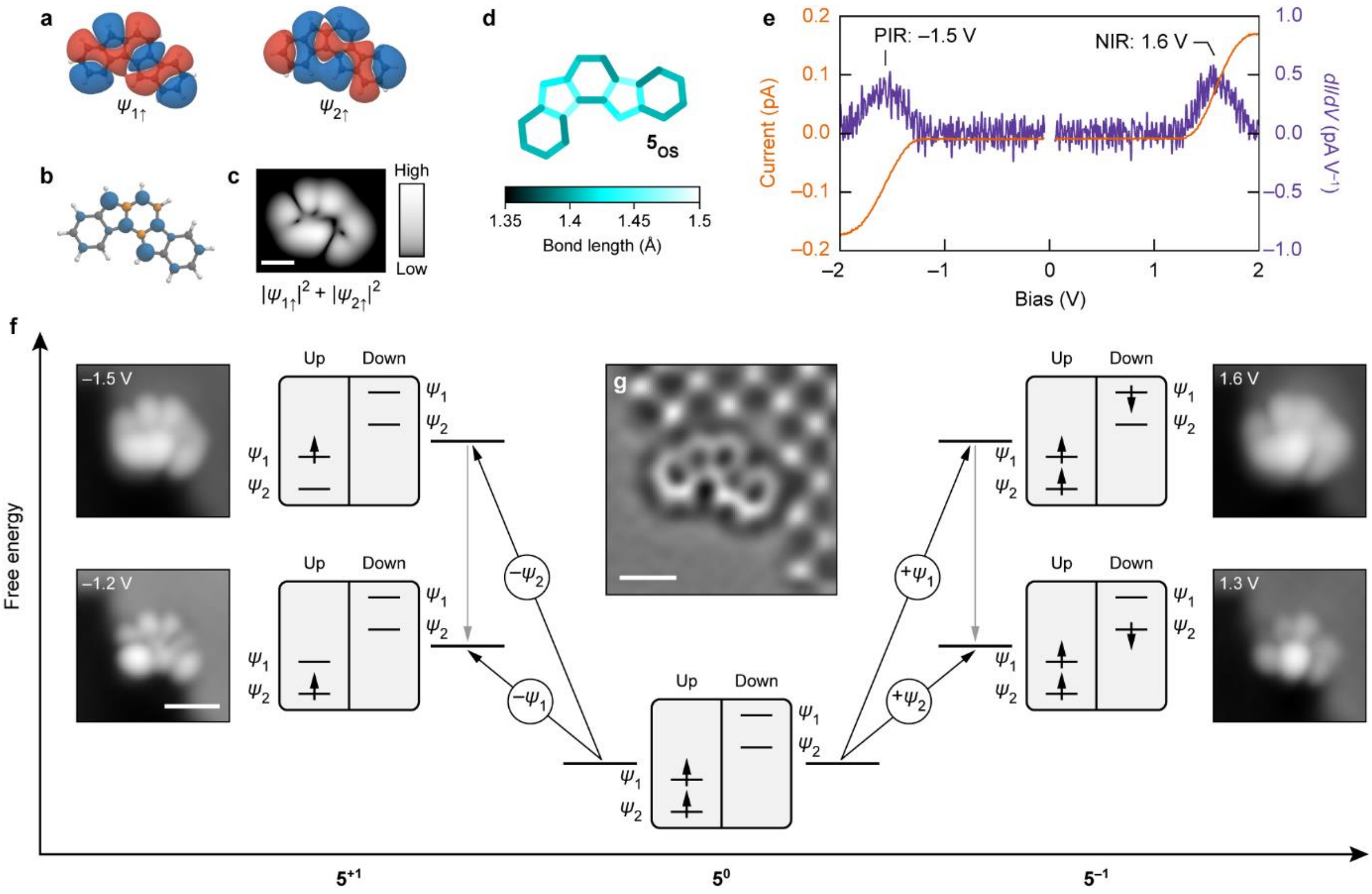

**Fig. 2 | Characterization of open-shell indeno[1,2-*a*]fluorene on bilayer NaCl/Au(111). a,** DFT-calculated wave functions of the frontier orbitals of $5_{OS}$ in the triplet configuration for the spin up (occupied) level (isovalue: 0.002 $e^{-}$ $Å^{-3}$). Blue and red colors represent opposite phases of the wave function. Orbital densities (wave functions squared) are shown in Supplementary Fig. 12. **b,** Corresponding DFT-calculated spin density of $5_{OS}$ (isovalue: 0.01 $e^{-}$ $Å^{-3}$). Blue and orange colors represent spin up and spin down densities, respectively. **c,** Mean-field Hubbard local density of states map of the superposition of the SOMOs of $5_{OS}$, calculated at a height of 7 Å above the molecular plane. **d,** DFT-calculated bond lengths of $5_{OS}$. **e,** Constant-height $I(V)$ spectra acquired on a species of **5** assigned as $5_{OS}$, along with the corresponding $dI/dV(V)$ spectra. Open feedback parameters: $V = –2$ V, $I = 0.17$ pA (negative bias side) and $V = 2$ V, $I = 0.17$ pA (positive bias side). Acquisition position of the spectra is shown in Supplementary Fig. 7. **f,** Scheme of many-body transitions associated to the measured ionic resonances of $5_{OS}$. Also shown are STM images of assigned $5_{OS}$ at biases where the corresponding transitions become accessible. Scanning parameters: $I = 0.3$ pA ($V = –1.2$ V and –1.5 V) and 0.2 pA ($V = 1.3$ V and 1.6 V). **g,** Laplace-filtered AFM image of assigned $5_{OS}$. STM set point: $V = 0.2$ V, $I = 0.5$ pA on bilayer NaCl, $\Delta z = –0.3$ Å. The tip-height offset $\Delta z$ for each panel is provided with respect to the STM setpoint, and positive (negative) values of $\Delta z$ denote tip approach (retraction) from the STM setpoint. **f** and **g** show the same molecule at the same adsorption site, which is next to a third layer NaCl island. The bright and dark features in the third layer NaCl island in **g** correspond to $Cl^{-}$ and $Na^{+}$ ions, respectively. Scale bars: 5 Å (**c**,**g**) and 10 Å (**f**).

**Generation and characterization of indeno[1,2-*a*]fluorene on insulating surfaces.** To experimentally explore the electronic structure of **5**, we used bilayer NaCl films on coinage metal surfaces to electronically decouple the molecule from the metal surfaces. Before presenting the experimental findings, we summarize the results of our theoretical calculations performed on **5** in the neutral charge state (denoted as $\mathbf{5}^0$). We start by performing DFT calculations on $\mathbf{5}^0$ in the gas phase. Geometry optimization performed at the spin-unrestricted UB3LYP/6-31G level of theory leads to one local minimum, $\mathbf{5}_{\mathbf{OS}}$, the geometry of which corresponds to the open-shell resonance structure of **5** (Fig. 1d, Fig. 2d and Supplementary Tables 1–7, the label OS denotes open-shell). The triplet electronic configuration of $\mathbf{5}_{\mathbf{OS}}$ is the lowest-energy state, with the open-shell singlet configuration 90 meV higher in energy. Geometry optimization performed at the restricted closed-shell RB3LYP/6-31G level reveals two local minima, $\mathbf{5}_{\boldsymbol{para}}$ and $\mathbf{5}_{\boldsymbol{ortho}}$, the geometries of which (Fig. 3b) exhibit bond length alternations in line with the presence of a *para*- or an *ortho*-QDM moiety, respectively, in the *as*-indacene core of the closed-shell resonance structures of **5** (Fig. 1d)[37]. Relative to $\mathbf{5}_{\mathbf{OS}}$ in the triplet configuration, $\mathbf{5}_{\boldsymbol{para}}$ and $\mathbf{5}_{\boldsymbol{ortho}}$ are 0.40 and 0.43 eV higher in energy, respectively. Additional DFT results are shown in Supplementary Fig. 5. To gain more accurate insights into the theoretical electronic structure of **5**, we performed multireference perturbation theory calculations (Supplementary Fig. 6) based on quasi-degenerate second-order *n*-electron valence state perturbation theory (QD-NEVPT2). In so far as the order of the ground and excited states are concerned, the results of QD-NEVPT2 calculations qualitatively match with DFT calculations. For $\mathbf{5}_{\mathbf{OS}}$, the triplet configuration remains the lowest-energy state, with the open-shell singlet configuration 60 meV higher in energy. The energy differences between the open- and closed-shell states are substantially reduced in QD-NEVPT2 calculations, with $\mathbf{5}_{\boldsymbol{para}}$ and $\mathbf{5}_{\boldsymbol{ortho}}$ only 0.11 and 0.21 eV higher in energy, respectively, compared to $\mathbf{5}_{\mathbf{OS}}$ in the triplet configuration. We also performed nucleus-independent chemical shift calculations to probe local aromaticity of **5** in the open- and closed-shell states. While $\mathbf{5}_{\mathbf{OS}}$ in the triplet configuration exhibits local aromaticity at the terminal benzenoid rings, $\mathbf{5}_{\mathbf{OS}}$ in the open-shell singlet configuration, $\mathbf{5}_{\boldsymbol{para}}$ and $\mathbf{5}_{\boldsymbol{ortho}}$ all display antiaromaticity (Supplementary Fig. 6).

The choice of the insulating surface determines the charge state of **5**: while **5** adopts neutral charge state on the high work function bilayer NaCl/Au(111) surface (irrespective of its open- or closed-shell state, Supplementary Fig. 7), **5** exhibits charge bistability between $\mathbf{5}^0$ and the anionic state $\mathbf{5}^{-1}$ on the lower work function bilayer NaCl/Ag(111) and Cu(111) surfaces (Supplementary Figs. 8 and 9). In the main text, we focus on the characterization of **5** on bilayer NaCl/Au(111). Characterization of charge bistable **5** is reported in Supplementary Figs. 10 and 11. We first describe experiments on **5** on bilayer NaCl/Au(111), where **5** exhibits a geometry corresponding to the calculated $\mathbf{5}_{\mathbf{OS}}$ geometry, and an open-shell electronic configuration. We compare the experimental data on this species to calculations on $\mathbf{5}_{\mathbf{OS}}$ with a triplet configuration, as theory predicts a triplet ground state for $\mathbf{5}_{\mathbf{OS}}$. For $\mathbf{5}_{\mathbf{OS}}$, the calculated frontier orbitals correspond to the SOMOs $\psi_1$ and $\psi_2$ (Fig. 2a–c and Supplementary Fig. 12), whose spin up levels are occupied and the spin down levels are empty. Figure 2d shows the DFT-calculated bond lengths of $\mathbf{5}_{\mathbf{OS}}$, where the two salient features, namely, the small difference in the bond lengths within each ring and the notably longer bond lengths in the pentagonal rings, agree with the open-shell resonance structure of **5** (Fig. 1d).

Figure 2g shows an AFM image of **5** adsorbed on bilayer NaCl/Au(111) that we assign as $\mathbf{5}_{\mathbf{OS}}$, where the bond-order differences qualitatively correspond to the calculated $\mathbf{5}_{\mathbf{OS}}$ geometry (discussed and compared to the closed-shell state below). Differential conductance spectra ($dI/dV(V)$, where $I$ and $V$ denote the tunneling current and bias voltage, respectively) acquired on assigned $\mathbf{5}_{\mathbf{OS}}$ exhibit two peaks centered at –1.5 V and 1.6 V (Fig. 2e), which we assign to the positive and negative ion resonances (PIR and NIR), respectively. Figure 2f shows the corresponding STM images acquired at the onset ($V$ = –1.2 V/1.3 V) and the peak ($V$ = –1.5 V/1.6 V) of the ionic resonances. To draw a correspondence between the STM images and the molecular orbital densities, we consider tunneling events as many-body electronic transitions between different charge states of $\mathbf{5}_{\mathbf{OS}}$ (Fig. 2f). Within this framework, the PIR corresponds to transitions between $\mathbf{5^0}$ and the cationic state $\mathbf{5^{+1}}$. At the onset of the PIR at –1.2 V, an electron can only be detached from the SOMO $\psi_1$ and the corresponding STM image at –1.2 V shows the orbital density of $\psi_1$. Increasing the bias to the peak of the PIR at –1.5 V, it becomes possible to also empty the SOMO $\psi_2$, such that the corresponding STM image shows the superposition of $\psi_1$ and $\psi_2$, that is, $|\psi_1|^2 + |\psi_2|^2$ (ref. [39]). Similarly, the NIR corresponds to transitions between $\mathbf{5^0}$ and $\mathbf{5^{-1}}$. At the NIR onset of 1.3 V, only electron attachment to $\psi_2$ is energetically possible. At 1.6 V, electron attachment to $\psi_1$ also becomes possible, and the corresponding STM image shows the superposition of $\psi_1$ and $\psi_2$. The observation of the orbital densities of SOMOs, and not the hybridized HOMO and LUMO, proves the open-shell ground state of assigned $\mathbf{5}_{\mathrm{OS}}$. Measurements of the monoradical species with a doublet ground state are shown in Supplementary Fig. 13.

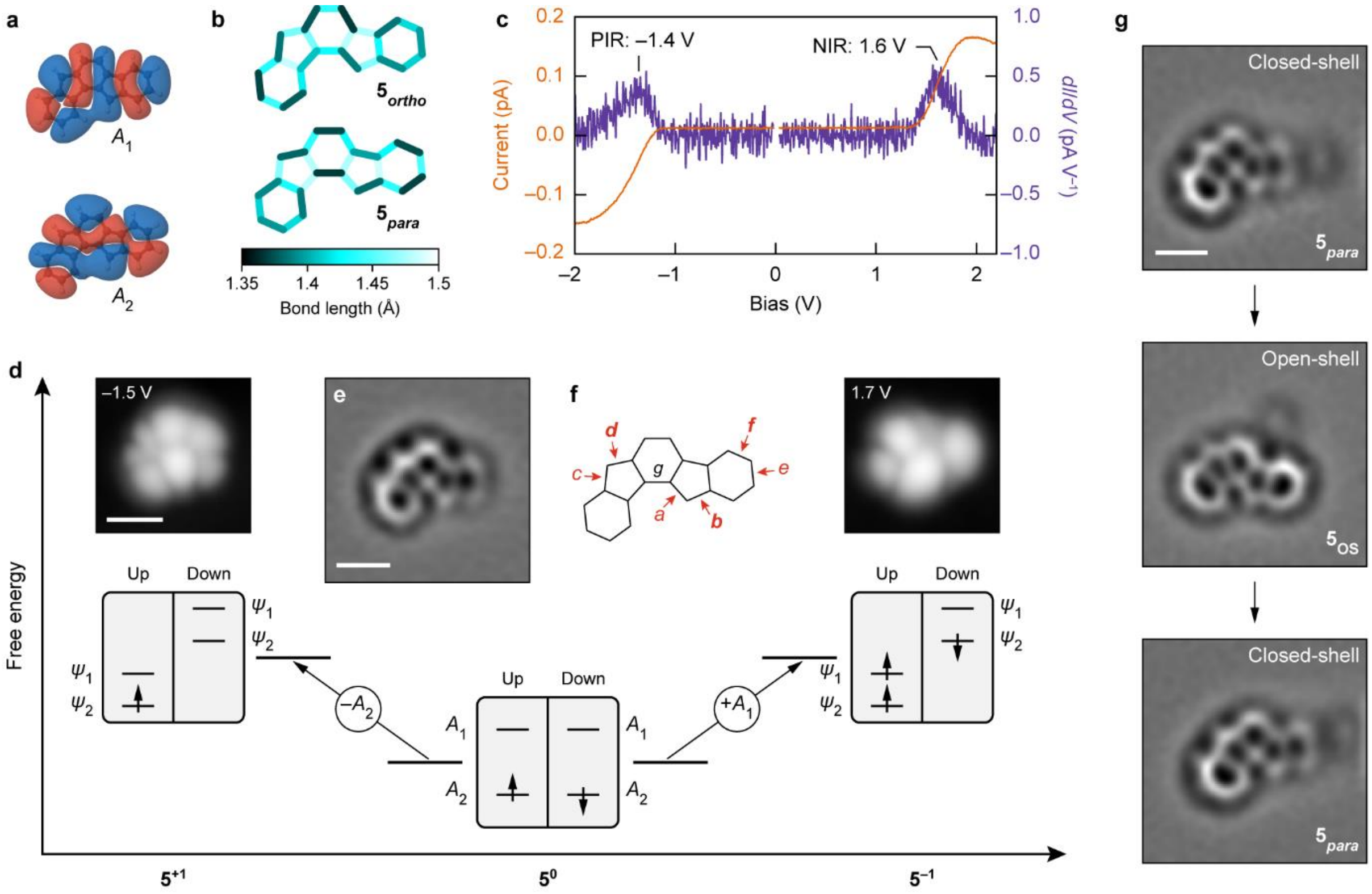

**Fig. 3 | Characterization of closed-shell indeno[1,2-*a*]fluorene on bilayer NaCl/Au(111). a,** DFT-calculated wave functions of the frontier orbitals of closed-shell $\mathbf{5^0}$ (isovalue: 0.002 e$^-$ Å$^{-3}$). The wave functions shown here are calculated for the $\mathbf{5}_{para}$ geometry. Orbital densities (wave functions squared) are shown in Supplementary Fig. 12. **b,** DFT-calculated bond lengths of $\mathbf{5}_{ortho}$ (top) and $\mathbf{5}_{para}$ (bottom). **c,** Constant-height $I(V)$ spectra acquired on a species of **5** assigned as $\mathbf{5}_{para}$, along with the corresponding $dI/dV(V)$ spectra. Open feedback parameters: $V$ = –2 V, $I$ = 0.15 pA (negative bias side) and $V$ = 2.2 V, $I$ = 0.15 pA (positive bias side). Acquisition position of the spectra is shown in Supplementary Fig. 7. **d,** Scheme of many-body transitions associated to the measured ionic resonances of $\mathbf{5}_{para}$. Also shown are STM images of assigned $\mathbf{5}_{para}$ at biases where the corresponding transitions become accessible. Scanning parameters: $I$ = 0.15 pA ($V$ = –1.5 V) and 0.2 pA ($V$ = 1.7 V). **e,** Laplace-filtered AFM image of assigned $\mathbf{5}_{para}$. STM set point: $V$ = 0.2 V, $I$ = 0.5 pA on bilayer NaCl, $\Delta z$ = –0.7 Å. Here, the molecule is adsorbed on top of a defect on the surface. For an example of a $\mathbf{5}_{para}$ species adsorbed adjacent to a third layer NaCl island, see Supplementary Fig. 15. **f,** Selected bonds labeled for highlighting bond order differences between $\mathbf{5}_{para}$ and $\mathbf{5}_{ortho}$. For the bond pairs *a*/*b*, *c*/*d* and *e*/*f*, the bonds labeled in bold exhibit a higher bond order than their neighboring labeled bonds in $\mathbf{5}_{para}$. **g,** Laplace-filtered AFM images of **5** on bilayer NaCl/Cu(111) showing switching between $\mathbf{5}_{\mathrm{OS}}$ and $\mathbf{5}_{para}$ as the molecule changes its adsorption position. Switching from $\mathbf{5}_{para}$ to $\mathbf{5}_{\mathrm{OS}}$ was induced by scanning

at 1.1 V, while switching from **5**$_{OS}$ back to **5**$_{para}$ took place by scanning at –2.2 V. The faint protrusion adjacent to **5** is a defect that stabilizes the adsorption of **5**. STM set point: $V$ = 0.2 V, $I$ = 0.5 pA on bilayer NaCl, $\Delta z$ = –0.3 Å. STM and STS data in **c** and **d** are acquired on the same species, while the AFM data in **e** is acquired on a different species. Scale bars: 10 Å (**d**) and 5 Å (**e,g**).

Unexpectedly, another species of **5** was also experimentally observed that exhibited a closed-shell ground state. In contrast to **5**$_{OS}$, where the frontier orbitals correspond to the SOMOs $\psi_1$ and $\psi_2$, DFT calculations predict orbitals of different shapes and symmetries for **5**$_{para}$ and **5**$_{ortho}$, denoted as $A_1$ and $A_2$ and shown in Fig. 3a and Supplementary Fig. 12. For **5**$_{ortho}$, $A_1$ and $A_2$ correspond to HOMO and LUMO, respectively. The orbitals are inverted in energy and occupation for **5**$_{para}$, where $A_2$ is the HOMO and $A_1$ is the LUMO. Figure 3e shows an AFM image of **5** that we assign as **5**$_{para}$. We experimentally infer its closed-shell state first by using qualitative bond order discrimination by AFM. In high-resolution AFM imaging, chemical bonds with higher bond order are imaged brighter (that is, with higher frequency shift $\Delta f$) due to stronger repulsive forces, and they appear shorter[34,47,48]. In Fig. 3f, we label seven bonds whose bond orders show significant qualitative differences in the calculated **5**$_{ortho}$, **5**$_{para}$ (Fig. 3b) and **5**$_{OS}$ (Fig. 2d) geometries. In **5**$_{para}$, the bonds *b* and *d* exhibit a higher bond order than *a* and *c*, respectively. This pattern is reversed for **5**$_{ortho}$, while the bond orders of the bonds *a*–*d* are all similar and small for **5**$_{OS}$. Furthermore, in **5**$_{para}$ bond *f* exhibits a higher bond order than *e*, while in **5**$_{ortho}$ and **5**$_{OS}$ bonds *e* and *f* exhibit similar bond order (because they belong to Clar sextets). Finally, the bond labeled *g* shows a higher bond order in **5**$_{para}$ than in **5**$_{ortho}$ and **5**$_{OS}$. The AFM image of assigned **5**$_{para}$ shown in Fig. 3e indicates higher bond orders of the bonds *b*, *d* and *f* compared to *a*, *c* and *e*, respectively. In addition, the bond *g* appears almost point-like and with enhanced $\Delta f$ contrast compared to its neighboring bonds, indicative of a high bond order (see Supplementary Fig. 14 for height-dependent measurements). These observations concur with the calculated **5**$_{para}$ geometry (Fig. 3b). Importantly, all these distinguishing bond-order differences are distinctly different in the AFM image of **5**$_{OS}$ shown in Fig. 2g, which is consistent with the calculated **5**$_{OS}$ geometry (Fig. 2d). In the AFM images of **5**$_{OS}$ (Fig. 2g and Supplementary Fig. 10), the bonds *a*–*d* at the pentagon apices appear with similar contrast and apparent bond length. The bonds *e* and *f* at one of the terminal benzenoid rings also exhibit similar contrast and apparent bond length, while the central bond *g* appears longer compared to assigned **5**$_{para}$.

Further compelling evidence for the closed-shell state of assigned **5**$_{para}$ is obtained by STM and STS. *dI/dV*(*V*) spectra acquired on an assigned **5**$_{para}$ species exhibit two peaks centered at –1.4 V (PIR) and 1.6 V (NIR) (Fig. 3c). STM images acquired at these biases (Fig. 3d) show the orbital densities of $A_2$ (PIR) and $A_1$ (NIR). First, the observation of $A_1$ and $A_2$ as the frontier orbitals of this species, and not the SOMOs, strongly indicates its closed-shell state. Second, consistent with AFM measurements that indicate good correspondence to the calculated **5**$_{para}$ geometry, we observe $A_2$ as the HOMO and $A_1$ as the LUMO. For **5**$_{ortho}$, $A_1$ should be observed as the HOMO and $A_2$ as the LUMO. We did not observe molecules with the signatures of **5**$_{ortho}$ in our experiments.

We observed molecules in open- (**5**$_{OS}$, Fig. 2) and closed-shell (**5**$_{para}$, Fig. 3) states in similar occurrence after their generation from **6** on the surface (out of 47 molecules, 23 and 24 molecules corresponded to **5**$_{OS}$ and **5**$_{para}$, respectively). We could also switch individual molecules between open- and closed-shell states as shown in Fig. 3g and Supplementary Fig. 15. To this end, a change in the adsorption site of a molecule (whether **5**$_{OS}$ or **5**$_{para}$) was induced by STM imaging at either of the ionic resonances, which often resulted in movement of the molecule. The example presented in Fig. 3g shows a molecule that was switched from **5**$_{para}$ to **5**$_{OS}$ and back to **5**$_{para}$. The switching is not directed, that is, we cannot choose which of the two species will be formed when changing the adsorption site. Out of 22 instances where the molecules moved, 14 resulted in switching between $\mathbf{5}_{\mathrm{OS}}$ and $\mathbf{5}_{para}$, while in 8 instances there was no switching of the ground state. Furthermore, we observed $\mathbf{5}_{\mathrm{OS}}$ and $\mathbf{5}_{para}$ in equal yield upon changing the adsorption site. The molecule in Fig. 3e is adsorbed on top of a defect that stabilizes its adsorption geometry on bilayer NaCl. At defect-free adsorption sites on bilayer NaCl, that is, without a third layer NaCl island or atomic defects in the vicinity of the molecule, **5** could be stably imaged neither by AFM nor by STM at ionic resonances (Supplementary Fig. 9). Without changing the adsorption site, the state of **5** (open- or closed-shell) never changed, including the experiments on bilayer NaCl/Ag(111) and Cu(111), on which the charge state of **5** could be switched (Supplementary Figs. 8 and 9). Also on these lower work function surfaces, both open- and closed-shell species were observed for $\mathbf{5}^{0}$ and both showed charge bistability[36] between $\mathbf{5}^{0}$ ($\mathbf{5}_{\mathrm{OS}}$ or $\mathbf{5}_{para}$) and $\mathbf{5}^{-1}$ (Supplementary Figs. 10 and 11). The geometrical structure of $\mathbf{5}^{-1}$ probed by AFM, and its electronic structure probed by STM imaging at the NIR (corresponding to transitions between $\mathbf{5}^{-1}$ and the dianionic state $\mathbf{5}^{-2}$), is identical within the measurement accuracy for the charged species of both $\mathbf{5}_{\mathrm{OS}}$ and $\mathbf{5}_{para}$. When cycling the charge state of **5** between $\mathbf{5}^{0}$ and $\mathbf{5}^{-1}$ several times, we always observed the same state ($\mathbf{5}_{\mathrm{OS}}$ or $\mathbf{5}_{para}$) when returning to $\mathbf{5}^{0}$, provided the molecule did not move during the charging/discharging process. For a discussion pertaining to the stabilization of and switching between the open- and closed-shell states of **5**, see Supplementary Note 1, and Supplementary Figs. 16 and 17.

## Conclusions

Based on our experimental observations we conclude that indeno[1,2-*a*]fluorene (**5**), the last unknown indenofluorene isomer, can be stabilized in and switched between an open-shell ($\mathbf{5}_{\mathrm{OS}}$) and a closed-shell ($\mathbf{5}_{para}$) state on NaCl. For the former, both DFT and QD-NEVPT2 calculations predict a triplet electronic configuration. Therefore, **5** can be considered to exhibit the spin-crossover effect, involving magnetic switching between high-spin ($\mathbf{5}_{\mathrm{OS}}$) and low-spin ($\mathbf{5}_{para}$) states, coupled with a reversible structural transformation. So far, the spin-crossover effect has mainly been observed in transition-metal-based coordination compounds with a near-octahedral geometry[49], with relatively few examples of polycyclic conjugated hydrocarbons exhibiting the effect[50]. The observation that the switching between open- and closed-shell states is related to changes in the adsorption site but is not achieved by charge-state cycling alone, indicates that the NaCl surface and local defects facilitate different electronic configurations of **5** depending on the adsorption site. Gas-phase QD-NEVPT2 calculations predict that $\mathbf{5}_{\mathrm{OS}}$ is the ground state, and the closed-shell $\mathbf{5}_{para}$ and $\mathbf{5}_{ortho}$ states are 0.11 and 0.21 eV higher in energy. The experiments, showing bidirectional switching between $\mathbf{5}_{\mathrm{OS}}$ and $\mathbf{5}_{para}$, indicate that a change in the adsorption site can induce sufficient change in the geometry of **5** (leading to a corresponding change in the ground state electronic configuration) and thus induce switching. Switching between open- and closed-shell states in **5** does not require the formation or dissociation of covalent bonds[51], but a change of adsorption site on NaCl where the molecule is physisorbed.

Our results should have implications for single-molecule devices, capitalizing on the altered electronic and chemical properties of a system in π-diradical open-shell and closed-shell states such as frontier orbital and singlet-triplet gaps, and chemical reactivity. For possible future applications as a single-molecule switch, it might be possible to also switch between open- and closed-shell states by changing the local electric field, such as by using chargeable adsorbates[52].

## Methods

**Scanning probe microscopy measurements and sample preparation.** STM and AFM measurements were performed in a home-built system operating at base pressures below $1\times10^{-10}$ mbar and a base temperature of 5 K. Bias voltages are provided with respect to the sample. All STM, AFM and spectroscopy measurements were performed with carbon monoxide (CO) functionalized tips. AFM measurements were performed in non-contact mode with a qPlus sensor[53]. The sensor was operated in frequency modulation mode[54] with a constant oscillation amplitude of 0.5 Å. STM measurements were performed in constant-current mode, AFM measurements were performed in constant-height mode with $V = 0$ V, and $I(V)$ and $\Delta f(V)$ spectra were acquired in constant-height mode. Positive (negative) values of the tip-height offset $\Delta z$ represent tip approach (retraction) from the STM setpoint. All $dI/dV(V)$ spectra are obtained by numerical differentiation of the corresponding $I(V)$ spectra. STM and AFM images, and spectroscopy curves, were post-processed using Gaussian low-pass filters.

Au(111), Ag(111) and Cu(111) surfaces were cleaned by iterative cycles of sputtering with $Ne^+$ ions and annealing up to 800 K. NaCl was thermally evaporated on Au(111), Ag(111) and Cu(111) surfaces held at 323 K, 303 K and 283 K, respectively. This protocol results in the growth of predominantly bilayer (100)-terminated islands, with a minority of third layer islands. Sub-monolayer coverage of **6** on the surfaces was obtained by flashing an oxidized silicon wafer containing the precursor molecules in front of the cold sample in the microscope. CO molecules for tip functionalization were dosed from the gas phase on the cold sample.

**Mean-field Hubbard calculations.** Tight-binding/mean-field Hubbard calculations have been performed by numerically solving the mean-field Hubbard Hamiltonian with nearest-neighbor hopping

$$\hat{H}_{MFH} = -t \sum_{\langle i,j\rangle,\sigma} c_{i,\sigma}^{\dagger} c_{j,\sigma} + U \sum_{i,\sigma} \langle n_{i,\sigma}\rangle n_{i,\bar{\sigma}} - U \sum_{i} \langle n_{i,\uparrow}\rangle\langle n_{i,\downarrow}\rangle. \quad (1)$$

Here, $c_{i,\sigma}^{\dagger}$ and $c_{j,\sigma}$ denote the spin selective ($\sigma \in \{\uparrow,\downarrow\}$ with $\bar{\sigma} \in \{\downarrow,\uparrow\}$) creation and annihilation operator at neighboring sites $i$ and $j$, $t$ = 2.7 eV is the nearest-neighbor hopping parameter, $U$ = 3.5 eV is the on-site Coulomb repulsion, $n_{i,\sigma}$ and $\langle n_{i,\sigma}\rangle$ denote the number operator and mean occupation number at site $i$, respectively. Orbital electron densities, $\rho$, of the $n^{\text{th}}$-eigenstate with energy $E_n$ have been simulated from the corresponding state vector $a_{n,k,\sigma}$ by

$$\rho_{n,\sigma}(\vec{r}) = \left| \sum_{k} a_{n,k,\sigma} \phi_{2p_z}(\vec{r} - \vec{r}_k) \right|^2, \quad (2)$$

where $k$ denotes the atomic site index, and $\phi_{2p_z}$ is the Slater $2p_z$ orbital for carbon.

**Density functional theory calculations.** Gas-phase DFT was employed using the PSI4 program package[55]. All molecules with different charge (neutral and anionic) and electronic (open- and closed-shell) states were independently investigated. The B3LYP exchange-correlation functional with 6-31G basis set was employed for structural relaxation and single-point energy calculations. The convergence criteria were set to $10^{-4}$ eV $Å^{-1}$ for the total forces and $10^{-6}$ eV for the total energies.

For the on-surface DFT calculations shown in Supplementary Fig. 16, we employed the FHI-aims[56] package. Molecules in the open- and closed-shell states were first independently investigated in the gas phase. The optimized molecular geometries were then optimized on a 9×9 bilayer NaCl slab in a cluster-type calculation. Molecular geometries in the gas phase were optimized with the *really tight* basis defaults. For the on-surface calculations we used light basis for NaCl atoms, and *really tight* basis defaults for atoms in the molecule. For structural relaxation, we employed the B3LYP exchange-correlation functional using the Vosko-Wilk-Nusair[57] local-density approximation, as implemented in the FHI-aims package.

In addition, we used the van der Waals scheme by Tkatchenko and Scheffler[58]. The convergence criteria for on-surface calculations were set to $10^{-3}$ eV Å$^{-1}$ for the total forces and $10^{-2}$ eV for the total energies. For the NaCl slab, we constrained the atoms at the edges of the slab, while the atoms located in the top NaCl layer away from the edges were allowed to relax.

We also studied **5** in the gas phase and adsorbed on bilayer NaCl in a 5×5 surface cell using periodic, plane-wave DFT calculations with VASP[59,60]. We employed the optB86b version of the van der Waals density functional[61–64], a plane-wave energy cutoff of 600 eV, and a 2×2 Monkhorst-Pack *k*-point mesh for the surface cell. The NaCl slab was constructed with a bulk lattice constant of 5.64 Å. Structural relaxations of molecule and top NaCl layer (with fixed bottom layer) were performed until residual forces were below $10^{-2}$ eV Å$^{-1}$. The VASP-calculated adsorption sites and their qualitative differences in energy for the open- and closed-shell states were found to be consistent with the other DFT calculations shown in the main text and the supplementary information.

**Multireference calculations.** Multireference calculations were performed on the DFT-optimized geometries using the QD-NEVPT2 level of theory[65,66], with three singlet roots and one triplet root included in the state-averaged calculation. A (10,10) active space (that is, 10 electrons in 10 orbitals) was used along with the def2-TZVP basis set[67]. Increasing either the active space size or expanding the basis set resulted in changes of about 50 meV for relative energies of the singlet and triplet states. These calculations were performed using the ORCA package[68].

**Nucleus-independent chemical shift (NICS) calculations.** Isotropic nucleus-independent chemical shift values were evaluated at the centre of each ring using the B3LYP exchange-correlation functional with def2-TZVP basis set using the Gaussian 16 software package[69].

## Data availability

Materials and methods, solution synthesis and characterization of **6**, additional STM and AFM data of **5**, STM and AFM data of monoradical species, analysis of $\Delta f(V)$ spectra, and additional calculations are available in the supplementary information. Output files of DFT and multireference calculations are available at https://doi.org/10.5281/zenodo.8234159.

## Acknowledgements

We thank Harry L. Anderson, Kristjan Eimre, Miquel Solà and Géza Giedke for fruitful discussions. This work was financially supported by the European Union project SPRING (grant number 863098), the European Research Council Synergy grant MolDAM (grant number 951519), the H2020-MSCA-ITN ULTIMATE (grant number 813036), the Spanish Agencia Estatal de Investigación (PID2019-107338RB-C62 and PID2020-115406GB-I00), Xunta de Galicia (Centro de Investigación de Galicia accreditation 2019–2022, ED431G 2019/03), and the European Regional Development Fund. This work used the Cirrus UK National Tier-2 high-performance computing service at EPCC (http://www.cirrus.ac.uk), funded by the University of Edinburgh and the Engineering and Physical Sciences Research Council (EP/P020267/1). I.R. acknowledges funding from the UK Research and Innovation (project ElDelPath, EP/X030075/1).

## Author contributions

S.M. and L.G. performed the on-surface synthesis and scanning probe microscopy measurements. M.V.-V. and D.P. synthesized and characterized the precursor molecule in solution. L.-A.L., S.F., I.R. and T.F. performed the density functional theory calculations. I.R. performed the NICS calculations. S.M., R.O. and T.F. performed the tight-binding calculations. R.O., I.R. and T.F. performed the multireference calculations. S.M. drafted the first version of the manuscript, and all authors contributed to discussing the results and writing the manuscript.

## Competing interests

The authors declare no competing interests.

## Additional information

**Correspondence and requests for materials** should be addressed to Shantanu Mishra, Diego Peña or Leo Gross.

# Supplementary Information

# Bistability between π-diradical open-shell and closed-shell states in indeno[1,2-*a*]fluorene

Shantanu Mishra[1], Manuel Vilas-Varela[2], Leonard-Alexander Lieske[1], Ricardo Ortiz[3], Shadi Fatayer[4], Igor Rončević[5], Florian Albrecht[1], Thomas Frederiksen[3,6], Diego Peña[2] and Leo Gross[1]

[1]IBM Research Europe – Zurich, 8803 Rüschlikon, Switzerland

[2]Center for Research in Biological Chemistry and Molecular Materials (CiQUS) and Department of Organic Chemistry, University of Santiago de Compostela, 15782 Santiago de Compostela, Spain

[3]Donostia International Physics Center (DIPC), 20018 Donostia-San Sebastián, Spain

[4]Physical Science and Engineering Division, King Abdullah University of Science and Technology (KAUST), 23955-6900 Thuwal, Saudi Arabia

[5]Department of Chemistry, Oxford University, Oxford OX1 3TA, United Kingdom

[6]Ikerbasque, Basque Foundation for Science, 48013 Bilbao, Spain

**Contents:**

## 1. Solution synthesis and characterization data.

### Experimental details

Starting materials (reagent grade) were purchased from TCI and Sigma-Aldrich and used without further purification. Reactions were carried out in flame-dried glassware and under an inert atmosphere of purified Ar using Schlenk techniques. Thin-layer chromatography (TLC) was performed on Silica Gel 60 F-254 plates (Merck). Column chromatography was performed on silica gel (40-60 µm). Nuclear magnetic resonance (NMR) spectra were recorded on a Bruker Varian Mercury 300 or Bruker Varian Inova 500 spectrometers. Mass spectrometry (MS) data were recorded in a Bruker Micro-TOF spectrometer. The synthesis of compound **6** was developed following the two-step synthetic route shown in Supplementary Fig. 1, which is based on the preparation of methylene-bridge polyarenes by means of Pd-catalyzed activation of benzylic C-H bonds[1].

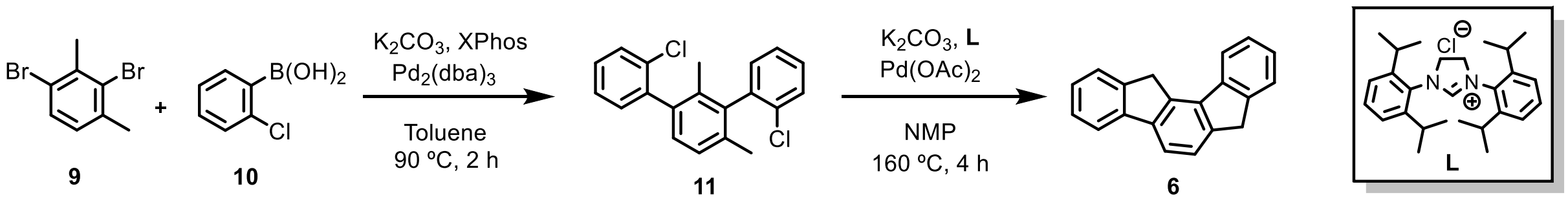

**Supplementary Fig. 1 |** Synthetic route to obtain compound **6**.

### Synthesis of 2,2''-dichloro-2',4'-dimethyl-1,1':3',1''-terphenyl (11)

The complex $Pd_2(dba)_3$ (20 mg, 0.02 mmol) was added over a deoxygenated mixture of 1,3-dibromo-2,4-dimethylbenzene (**9**, 100 mg, 0.38 mmol), boronic acid **10** (178 mg, 1.14 mmol), $K_2CO_3$ (314 mg, 2.28 mmol) and XPhos (35 mg, 0.08 mmol) in toluene (1:1, 10 mL), and the resulting mixture was heated at 90 °C for 2 h. After cooling to room temperature, the solvents were evaporated under reduced pressure. The reaction crude was purified by column chromatography ($SiO_2$; hexane:$CH_2Cl_2$ 9:1) affording **11** (94 mg, 76%) as a colorless oil. **$^1$H NMR** (300 MHz, $CDCl_3$) δ: 7.51 (m, 2H), 7.40 – 7.28 (m, 5H), 7.27 – 7.20 (m, 2H), 7.13 (d, *J* = 7.7 Hz, 1H), 2.07 (s, 3H), 1.77 (s, 3H) ppm. **$^{13}$C NMR-DEPT** (75 MHz, $CDCl_3$, 1:1 mixture of atropisomers) δ: 141.2 (C), 141.1 (C), 140.0 (C), 139.4 (2C), 137.5 (C), 137.4 (C), 136.0 (3C), 134.8 (C), 134.5 (C), 134.1 (C), 134.0 (C), 133.7 (C), 133.6 (C), 131.6 (CH), 131.2 (CH), 131.1 (CH), 130.7 (CH), 129.8 (CH), 129.7 (CH), 129.5 (CH), 129.4 (CH), 129.0 (CH), 128.9 (CH), 128.7 (2CH), 128.6 (2CH), 127.2 (CH), 127.1 (CH), 127.0 (CH), 126.9 (CH), 126.7 (CH), 126.6 (CH), 20.6 ($CH_3$), 20.5 ($CH_3$), 17.7 ($CH_3$), 17.5 ($CH_3$) ppm. **MS (APCI)** *m/z* (%): 327 (M+1, 100). **HRMS**: $C_{20}H_{16}Cl_2$; calculated: 327.0702, found: 327.0709.

### Synthesis of 7,12-dihydroindeno[1,2–*a*]fluorene (6)

The complex $Pd(OAc)_2$ (7 mg, 0.03 mmol) was added over a deoxygenated mixture of terphenyl **11** (90 mg, 0.27 mmol), $K_2CO_3$ (114 mg, 0.83 mmol) and ligand **L** (26 mg, 0.06 mmol) in NMP (2 mL). The resulting mixture was heated at 160 °C for 4 h. After cooling to room temperature, $H_2O$ (30 mL) was added, and the mixture was extracted with EtOAc (3x15 mL). The combined organic extracts were dried over anhydrous $Na_2SO_4$, filtered, and evaporated under reduced pressure. The reaction crude was purified by column chromatography ($SiO_2$; hexane:$CH_2Cl_2$ 9:1) affording compound **6** (8 mg, 11%) as a white solid. **$^1$H NMR** (500 MHz, $CDCl_3$) δ: 7.93 (d, *J* = 7.6 Hz, 1H), 7.85 (d, *J* = 7.5 Hz, 1H), 7.78 (d, *J* = 7.7 Hz, 1H), 7.65 (d, *J* = 7.4 Hz, 1H), 7.61 (d, *J* = 7.5 Hz, 1H), 7.59 (d, *J* = 7.7 Hz, 1H), 7.47 (ddd, *J* = 8.4, 7.2, 1.1 Hz, 1H), 7.42 (dd, J = 8.1, 7.0 Hz, 1H), 7.35 (m, 2H), 4.22 (s, 3H), 4.02 (s, 3H). ppm. **$^{13}$C NMR-DEPT** (125 MHz, $CDCl_3$) δ: 144.1 (C), 143.3 (C), 142.3 (C), 141.9 (C), 141.8 (C), 141.2 (C), 138.2 (C), 136.5 (C), 127.0 (CH), 126.9 (CH), 126.7 (CH), 126.6 (CH), 125.3 (CH), 125.2 (CH), 123.6 (CH), 122.2 (CH), 119.9 (CH), 118.4 (CH), 37.4 ($CH_2$), 36.3 ($CH_2$). ppm. **MS (APCI)** *m/z* (%): 254 (M+, 88). **HRMS**: $C_{20}H_{14}$; calculated: 254.1090, found: 254.1090.

## $^{1}H$ and $^{13}C$ NMR spectra

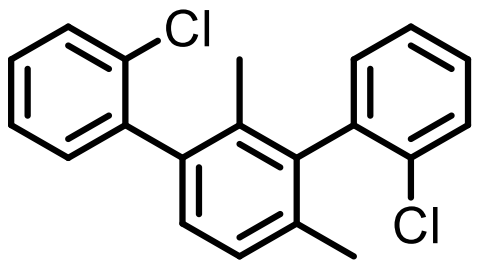

**11**

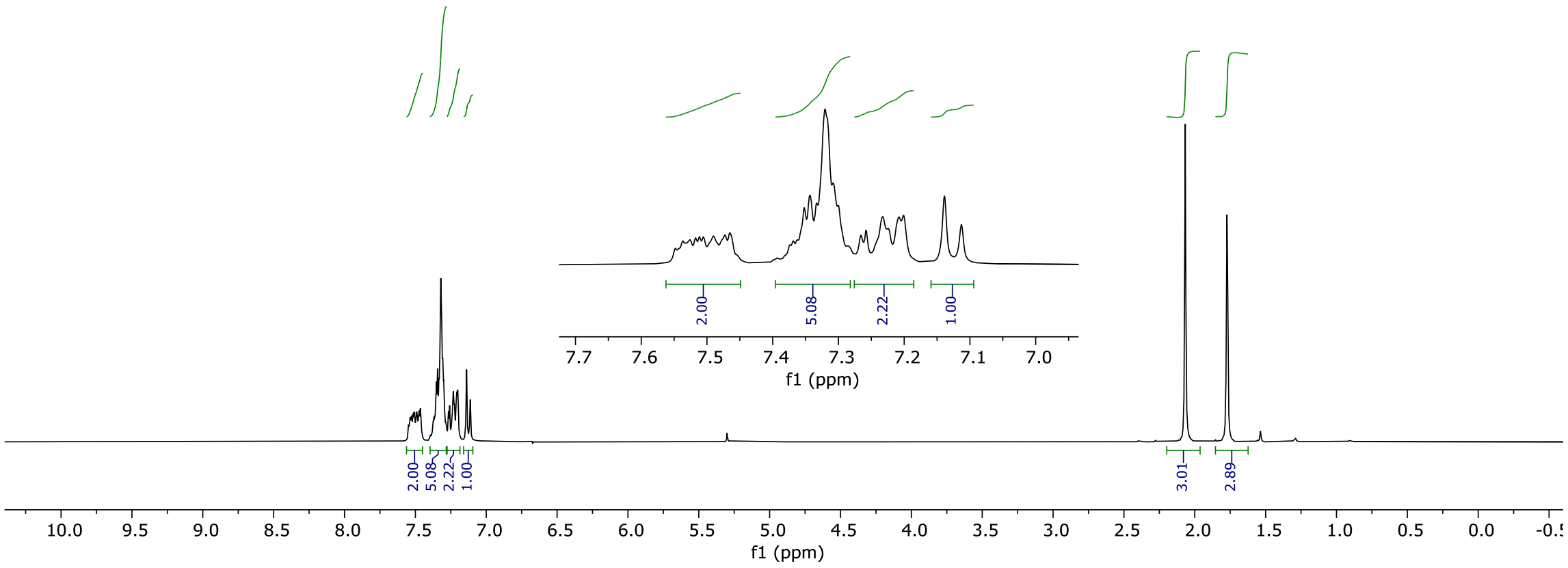

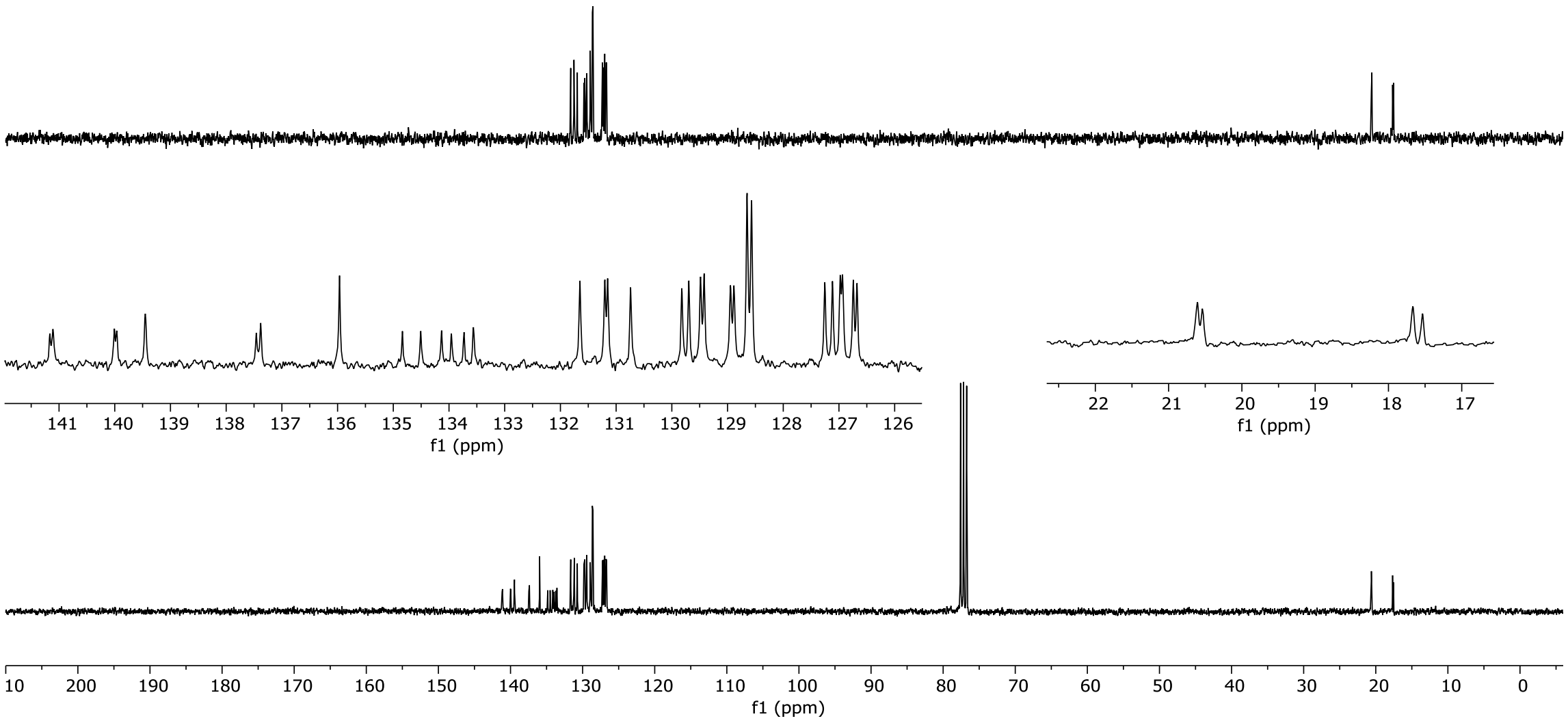

**Supplementary Fig. 2 |** $^{1}H$ (top) and $^{13}C$ NMR (bottom) spectra of compound **11**.

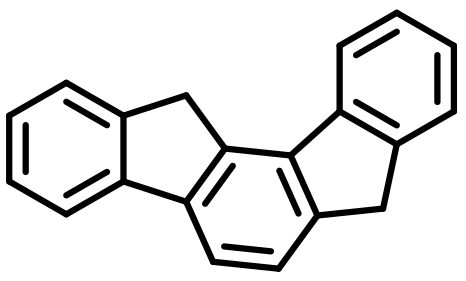

**6**

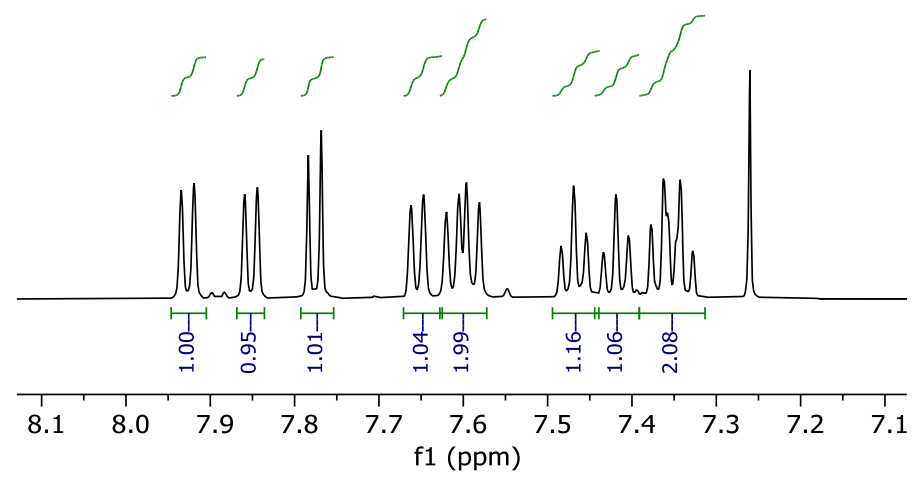
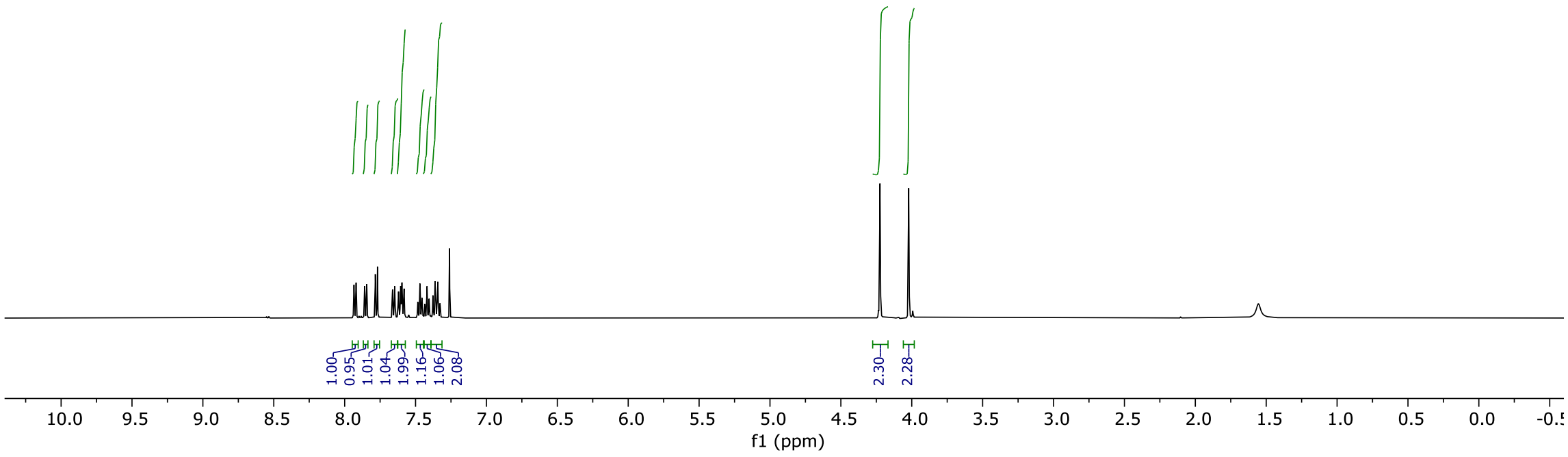

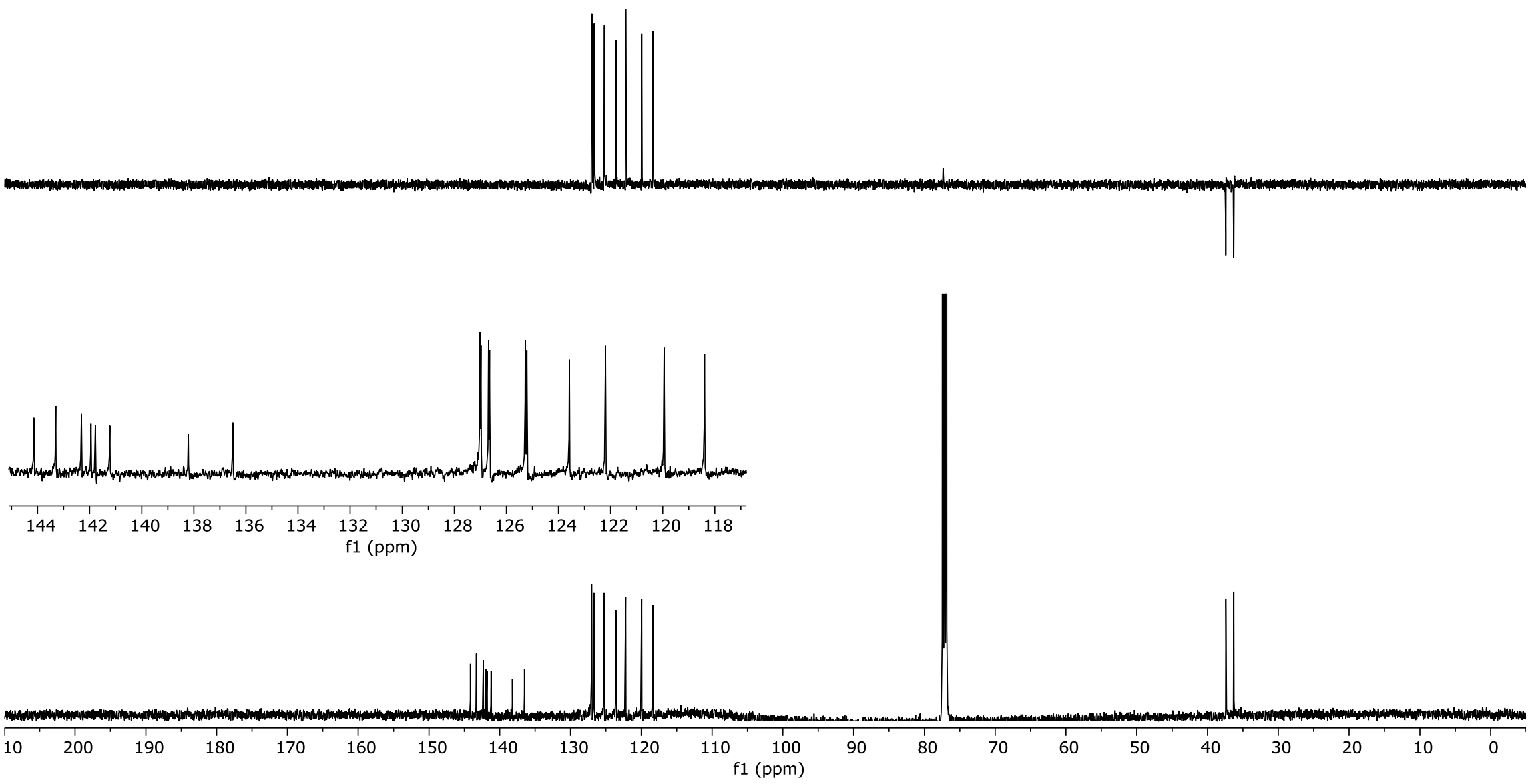

**Supplementary Fig. 3 |** $^{1}$H (top) and $^{13}$C (bottom) NMR spectra of compound **6**.

## 2. Scanning probe microscopy and spectroscopy data, and theoretical calculations.

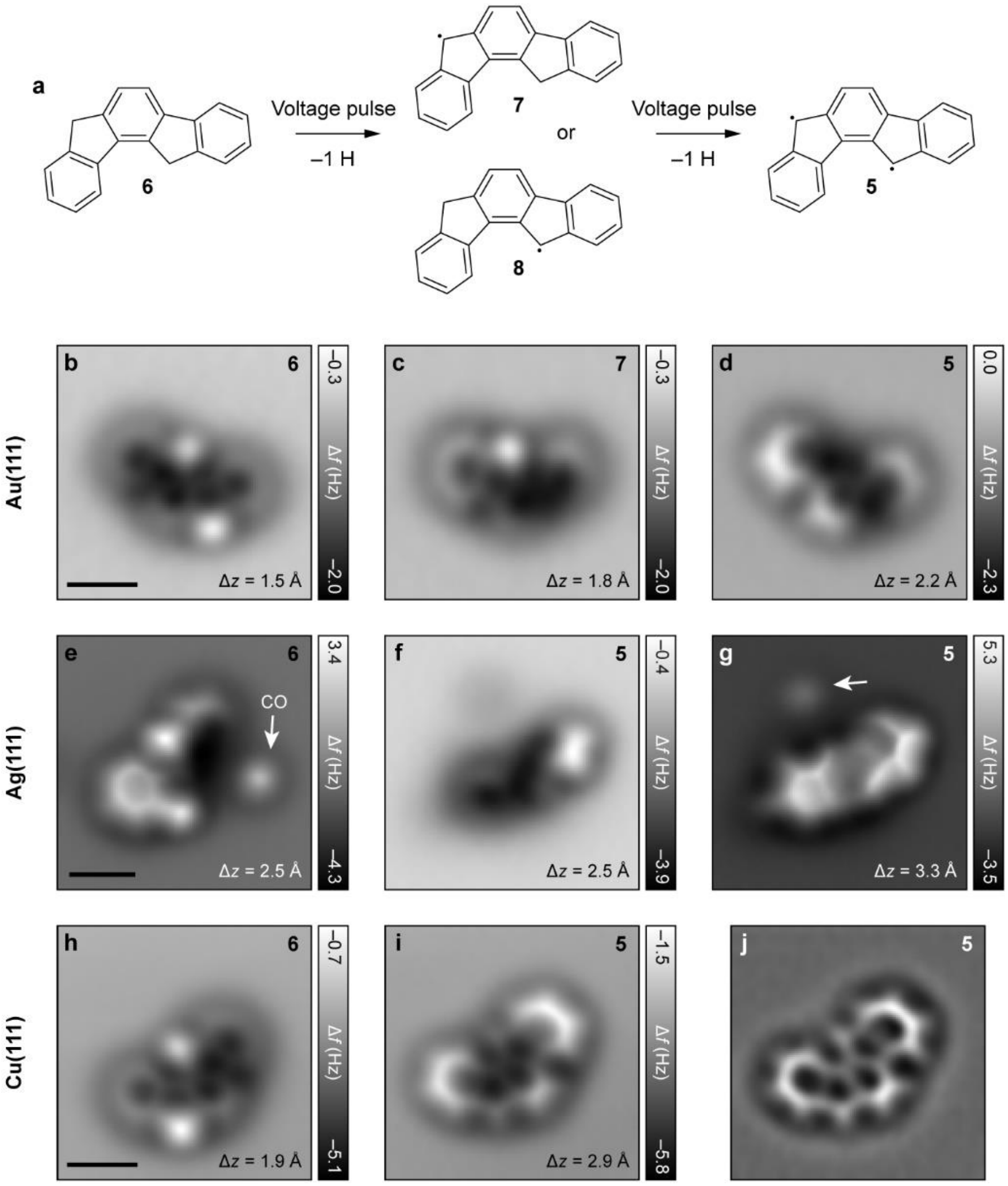

**Supplementary Fig. 4 | Generation of indeno[1,2-*a*]fluorene on coinage metal surfaces. a,** Scheme of on-surface generation of **5** ($C_{20}H_{12}$) by voltage pulse-induced dehydrogenation of **6** ($C_{20}H_{14}$). Structures **7** and **8** represent the two monoradical species ($C_{20}H_{13}$). To generate **5**, the STM tip was positioned at the center of **6** at typical tunneling conditions of $V$ = 0.2 V and $I$ = 1 pA, and the feedback loop was switched off to maintain a constant tip height relative to the molecule. The tip was then retracted by 5–10 Å to limit the tunneling current, and the bias was ramped to 4–6 V. Abrupt changes in the $I(V)$ spectra indicated manipulation events, and the area was subsequently imaged to monitor the occurrence of dehydrogenation. **b–d,** AFM images showing stepwise generation of **5** on Au(111). AFM images are shown for precursor **6** (**b**), in which the two bright protrusions correspond to -$CH_2$- moieties at the pentagonal ring apexes, **7** (**c**), and **5** (**d**). **e–g,** AFM images showing generation of **5** on Ag(111). The arrow in **e** indicates a CO molecule adjacent to **6**, while the arrow in **g** indicates an adsorbate adjacent to **5**. **f** and **g** show AFM images of **5** at two different heights. The tip is 0.8 Å closer to **5** in **g** than in **f**. **h,i,** AFM images showing generation of **5** on Cu(111). **j**, Laplace-filtered version of **i** revealing the carbon skeleton of **5**. On Cu(111), **5** adsorbs in a largely planar geometry, although the increased frequency shift in AFM imaging at the terminal benzenoid rings (leading to their brighter contrast) compared to the central rings implies a slight out-of-plane distortion of **5**. This feature was also observed for pentacene[2] and **2** (ref. [3]) on Cu(111). Additionally, qualitatively similar intramolecular resolution of **5** on Cu(111) is obtained at a much smaller tip-sample distance than on **6** (compare Δ$z$ values in **h** and **i**), indicating a substantially reduced adsorption height of **5**. On Au(111) (**d**) and Ag(111) (**g**), **5** adsorbs

in a non-planar conformation, where the apical carbon atoms of the pentagonal rings are not resolved in AFM imaging due to their reduced adsorption height compared to the rest of the carbon atoms. We attribute this observation to the significantly different lattice parameter of Cu(111) (2.57 Å) compared to Au(111) and Ag(111) (2.95 Å and 2.94 Å, respectively)[4], such that the apical carbon atoms of the pentagonal rings of **5** adsorb on the on-top atomic sites on Au(111) and Ag(111), but not on Cu(111). Our speculation is based on a previous study of polymers of **1** on Au(111) by Di Giovannantonio et al.[5], where both tilted and planar individual units of **1** were observed depending on whether the apical carbon atoms of the pentagonal rings in **1** adsorbed on the on-top or hollow sites of the surface, respectively. Given the strong molecule-metal interaction, we found no electronic state signatures of **5** on all three metal surfaces. STM set point for AFM images: *V* = 0.2 V and *I* = 0.5 pA on Au(111) and Ag(111), and *V* = 0.2 V and *I* = 1 pA on Cu(111). Scale bars: 5 Å.

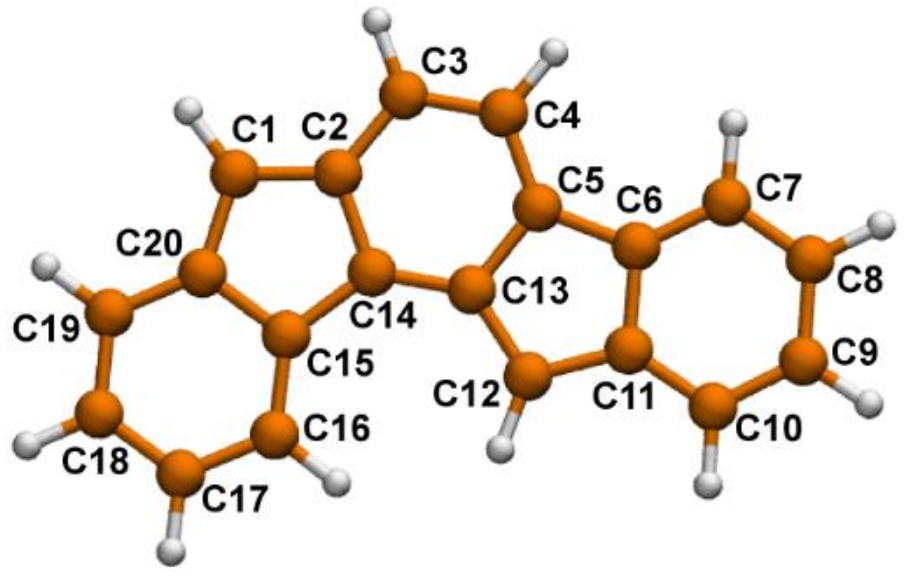

| Bond | Bond length, $\mathbf{5}_{\mathrm{OS}}$ | Bond length, $\mathbf{5}_{para}$ | Bond length, $\mathbf{5}_{ortho}$ | Bond length, $\mathbf{5}^{-1}$ |
|---|---|---|---|---|
| C1–C2 | 1.423 | 1.380 | 1.459 | 1.434 |
| C2–C3 | 1.412 | 1.438 | 1.376 | 1.417 |
| C3–C4 | 1.404 | 1.372 | 1.439 | 1.402 |
| C4–C5 | 1.394 | 1.429 | 1.368 | 1.398 |
| C5–C6 | 1.465 | 1.398 | 1.477 | 1.450 |
| C6–C7 | 1.392 | 1.426 | 1.385 | 1.398 |
| C7–C8 | 1.406 | 1.374 | 1.414 | 1.403 |
| C8–C9 | 1.405 | 1.446 | 1.396 | 1.410 |
| C9–C10 | 1.400 | 1.370 | 1.412 | 1.400 |
| C10–C11 | 1.403 | 1.437 | 1.392 | 1.410 |
| C11–C12 | 1.439 | 1.382 | 1.464 | 1.437 |
| C12–C13 | 1.428 | 1.461 | 1.387 | 1.404 |
| C13–C14 | 1.401 | 1.368 | 1.429 | 1.422 |
| C14–C15 | 1.466 | 1.481 | 1.407 | 1.444 |
| C15–C16 | 1.393 | 1.386 | 1.422 | 1.405 |
| C16–C17 | 1.407 | 1.415 | 1.378 | 1.399 |
| C17–C18 | 1.404 | 1.394 | 1.441 | 1.415 |
| C18–C19 | 1.401 | 1.412 | 1.371 | 1.395 |
| C19–C20 | 1.402 | 1.390 | 1.434 | 1.411 |
| C1–C20 | 1.441 | 1.466 | 1.384 | 1.423 |
| C2–C14 | 1.450 | 1.479 | 1.461 | 1.439 |
| C5–C13 | 1.456 | 1.476 | 1.480 | 1.480 |
| C6–C11 | 1.439 | 1.486 | 1.429 | 1.450 |
| C15–C20 | 1.441 | 1.430 | 1.483 | 1.460 |

**Supplementary Table 1 | DFT-calculated C–C bond lengths (in Å) of $\mathbf{5}_{\mathrm{OS}}$, $\mathbf{5}_{para}$, $\mathbf{5}_{ortho}$ and $\mathbf{5}^{-1}$.** Numbering scheme of carbon atoms is shown in the molecular model.

| Atom | X (Å) | Y (Å) | Z (Å) |
|---|---|---|---|
| C | -4.493874905187 | 1.674619220405 | -1.609303916742 |
| C | -3.863487509338 | 0.528779745012 | -2.119253744366 |
| C | -2.611822561254 | 0.113827348963 | -1.629441264251 |
| C | -1.996102060038 | 0.856210218342 | -0.624059854977 |
| C | -2.642110148231 | 2.032366121432 | -0.099920908974 |
| C | -3.889266818531 | 2.433231576145 | -0.598135899742 |
| C | -0.725251505360 | 0.704435280938 | 0.090931506587 |
| C | -0.629327165027 | 1.798198500825 | 1.037529797146 |
| C | -1.796805091569 | 2.602862599527 | 0.917952130093 |
| C | 0.318117383988 | -0.228299667792 | 0.025262767673 |
| C | 1.461241767848 | -0.063930174510 | 0.911432158857 |
| C | 1.536593440951 | 0.994329799145 | 1.815800584807 |
| C | 0.492527175395 | 1.930779898123 | 1.883999318271 |
| C | 0.535544659889 | -1.393262988491 | -0.772313363113 |
| C | 1.801567647606 | -1.970428907234 | -0.404050534403 |
| C | 2.387392905819 | -1.164000839683 | 0.633372852920 |
| C | 2.475643712545 | -3.106613430179 | -0.877416915878 |
| C | 3.719207610492 | -3.437170915929 | -0.324772042620 |
| C | 4.288121828211 | -2.646978444783 | 0.688394153431 |
| C | 3.624504160418 | -1.505646622364 | 1.172808493764 |
| H | -5.459573539461 | 1.975093163789 | -2.003101175612 |
| H | -4.347732620325 | -0.045753454284 | -2.902426535897 |
| H | -2.144054134383 | -0.774996617744 | -2.038999819066 |
| H | -4.381361793425 | 3.318822570739 | -0.207227631455 |
| H | -2.015611303951 | 3.493449520313 | 1.493020528289 |
| H | 2.397744598582 | 1.099699275791 | 2.468472126088 |
| H | 0.547513427309 | 2.756240089197 | 2.586921973707 |
| H | -0.129763282138 | -1.786484482106 | -1.528387260264 |
| H | 2.041280122519 | -3.721272006993 | -1.660127876839 |
| H | 4.250916952706 | -4.313266511090 | -0.681905848943 |
| H | 5.252996532539 | -2.921178864985 | 1.102875782251 |
| H | 4.077963948155 | -0.905768559282 | 1.956172095786 |

**Supplementary Table 2 | DFT-optimized gas phase geometry of $5_{OS}$.**

| Atom | X (Å) | Y (Å) | Z (Å) |
|---|---|---|---|
| C | -4.134416460352 | 2.814309431291 | -0.845551534497 |
| C | -3.640407266867 | 1.809607147554 | -1.752460841137 |
| C | -2.510360497149 | 1.072189732277 | -1.473391689028 |
| C | -1.806851583881 | 1.303250559133 | -0.258975697678 |
| C | -2.312412915420 | 2.336692786536 | 0.677048697414 |
| C | -3.493359800352 | 3.076034253535 | 0.338216034376 |
| C | -0.649964490800 | 0.731255759410 | 0.302591894163 |
| C | -0.428804510017 | 1.399481290028 | 1.582451710616 |
| C | -1.480453252946 | 2.390301188423 | 1.782001753306 |
| C | 0.269037218441 | -0.286494228615 | -0.100827307874 |
| C | 1.395180754790 | -0.604568725691 | 0.805767507828 |
| C | 1.566313962264 | 0.045720529979 | 1.997000128817 |
| C | 0.633773775906 | 1.068723114999 | 2.391684629095 |
| C | 0.373841369116 | -1.109111813579 | -1.212254070458 |
| C | 1.545718854217 | -1.975687378697 | -1.076443038658 |
| C | 2.185713159133 | -1.674768742806 | 0.165113205293 |
| C | 2.069992548441 | -2.957983016218 | -1.912297148076 |
| C | 3.236621389901 | -3.643427299416 | -1.509831233519 |
| C | 3.859342671754 | -3.347201104010 | -0.296602843074 |
| C | 3.330548845840 | -2.350260448835 | 0.555771330262 |
| H | -5.029409002597 | 3.365264701563 | -1.116443306242 |
| H | -4.180866951831 | 1.636844384159 | -2.678246526949 |
| H | -2.162794902644 | 0.322354134391 | -2.176237653349 |
| H | -3.867777378661 | 3.833860524632 | 1.020069392113 |
| H | -1.576416780992 | 3.038965035250 | 2.642011015389 |
| H | 2.395531896427 | -0.195100598260 | 2.654599156591 |
| H | 0.782863264368 | 1.576348743889 | 3.340669582436 |
| H | -0.295003555014 | -1.126747683052 | -2.062164576745 |
| H | 1.595429620084 | -3.198218041960 | -2.858713191362 |
| H | 3.652876510392 | -4.410917009506 | -2.154747534109 |
| H | 4.755412092791 | -3.883428054063 | -0.002166263143 |
| H | 3.822353475338 | -2.126845898377 | 1.498047369205 |

**Supplementary Table 3 | DFT-optimized gas phase geometry of $5_{ortho}$.**

| Atom | X (Å) | Y (Å) | Z (Å) |
|---|---|---|---|
| C | -3.389276477755 | 3.585428848613 | 1.027178952043 |
| C | -2.017911766127 | 3.761323537237 | 1.208339511802 |
| C | -1.109850316602 | 2.720208485429 | 0.903602820347 |
| C | -1.604865259809 | 1.518734727532 | 0.420707071205 |
| C | -3.011184710328 | 1.339298501435 | 0.235274102316 |
| C | -3.900655192096 | 2.364535838239 | 0.535767492019 |
| C | -0.954510753441 | 0.252430661494 | 0.011436525006 |
| C | -2.029089753515 | -0.668326057043 | -0.417758966204 |
| C | -3.236128110607 | -0.014627973278 | -0.280584333231 |
| C | 0.346664182123 | -0.166335624388 | -0.031541925290 |
| C | 0.638055604807 | -1.532834663686 | -0.508100210036 |
| C | -0.413615402248 | -2.408615945404 | -0.918844831197 |
| C | -1.721665159283 | -1.995607808125 | -0.878565605522 |
| C | 1.611807843022 | 0.478726096524 | 0.312714580776 |
| C | 2.620301782542 | -0.433034686700 | 0.063970396618 |
| C | 2.025852534361 | -1.694234541302 | -0.449290043167 |
| C | 4.046019322961 | -0.346354975939 | 0.217019143185 |
| C | 4.822235481152 | -1.425350465747 | -0.112831769245 |
| C | 4.239499962828 | -2.650729855385 | -0.611951360678 |
| C | 2.882498483024 | -2.786382815714 | -0.777121632266 |
| H | -4.072975836230 | 4.393796135899 | 1.265607288704 |
| H | -1.638949024555 | 4.704909994818 | 1.586752830020 |
| H | -0.046992027060 | 2.875089158012 | 1.050818837036 |
| H | -4.969695701787 | 2.235705394866 | 0.397584642917 |
| H | -4.210599358659 | -0.425823864268 | -0.514319373744 |
| H | -0.168315192456 | -3.408409956239 | -1.264787561340 |
| H | -2.527393729101 | -2.652458007972 | -1.188074530586 |
| H | 1.732136836821 | 1.482774972453 | 0.691313236552 |
| H | 4.493112322774 | 0.569677068469 | 0.591114355720 |
| H | 5.900946359438 | -1.374046695575 | -0.002276088703 |
| H | 4.901228762556 | -3.475252948843 | -0.858178735430 |
| H | 2.462791307205 | -3.714456719140 | -1.153306535775 |

**Supplementary Table 4 | DFT-optimized gas phase geometry of 5$_{\textit{para}}$.**

| Atom | X (Å) | Y (Å) | Z (Å) |
|---|---|---|---|
| C | -4.496738578962 | 1.708824260175 | -1.625653354891 |
| C | -3.860788175699 | 0.552835298733 | -2.136595005000 |
| C | -2.619380304249 | 0.131698446179 | -1.646823114410 |
| C | -1.986038554152 | 0.864630303977 | -0.628974113388 |
| C | -2.641794262136 | 2.057311186331 | -0.101552006701 |
| C | -3.893117894423 | 2.457441931798 | -0.615010167601 |
| C | -0.735996557671 | 0.699601012078 | 0.074202202667 |
| C | -0.637138866908 | 1.781767884751 | 1.016869378518 |
| C | -1.803496285863 | 2.608932409589 | 0.906708726568 |
| C | 0.308082418908 | -0.262273057992 | -0.006163081082 |
| C | 1.458897501728 | -0.084477836306 | 0.907924472745 |
| C | 1.522072131459 | 0.980185861855 | 1.812501362788 |
| C | 0.481585604013 | 1.917253649910 | 1.875215511143 |
| C | 0.536566551744 | -1.409580941133 | -0.782787607119 |
| C | 1.798088142965 | -1.981235777145 | -0.401323854729 |
| C | 2.384326459236 | -1.170026293088 | 0.648087547418 |
| C | 2.503603915808 | -3.116854258233 | -0.848918209301 |
| C | 3.741497819863 | -3.436105292764 | -0.278935700341 |
| C | 4.306091133464 | -2.643244699273 | 0.741125191822 |
| C | 3.621844781317 | -1.508181538623 | 1.202477941900 |
| H | -5.461516556915 | 2.012257346928 | -2.025584355764 |
| H | -4.347321051188 | -0.017937842520 | -2.924724031879 |
| H | -2.150661305504 | -0.758834906002 | -2.054583141165 |
| H | -4.384184971344 | 3.346668950402 | -0.223180959028 |
| H | -2.007108641425 | 3.498870929015 | 1.491568984299 |
| H | 2.382142148672 | 1.083088794982 | 2.471482458596 |
| H | 0.531696235301 | 2.745005116626 | 2.579317754140 |
| H | -0.117227905498 | -1.810899782963 | -1.545558825564 |
| H | 2.088913476297 | -3.744662625798 | -1.635019178226 |
| H | 4.279923557751 | -4.314780664127 | -0.630261857185 |
| H | 5.269103724923 | -2.911691323034 | 1.168251145415 |
| H | 4.057135025593 | -0.895278424907 | 1.990248785612 |

**Supplementary Table 5 | DFT-optimized gas phase geometry of $5^{-1}$.**

| Atom | X (Å) | Y (Å) | Z (Å) |
|---|---|---|---|
| C | -4.491439050731 | 1.686733076122 | -1.602351083526 |
| C | -3.841797181294 | 0.563790955315 | -2.139387665174 |
| C | -2.590597686532 | 0.147470862289 | -1.646338756751 |
| C | -1.997928821386 | 0.865769800635 | -0.612182179376 |
| C | -2.661250902060 | 2.018049494648 | -0.058884345566 |
| C | -3.908595381434 | 2.420326640321 | -0.561493874636 |
| C | -0.730243124316 | 0.703438641421 | 0.114531537399 |
| C | -0.656750745004 | 1.766301987286 | 1.094523200853 |
| C | -1.835580426518 | 2.564732651717 | 0.984341833637 |
| C | 0.306786430404 | -0.206852900546 | 0.029205368409 |
| C | 1.426912711730 | -0.081650146050 | 0.906497106243 |
| C | 1.498394164817 | 0.949651289874 | 1.857524840894 |
| C | 0.456246192539 | 1.875187605716 | 1.952733192436 |
| C | 0.498236847232 | -1.403705046747 | -0.886169970705 |
| C | 1.844875798271 | -1.953532887941 | -0.444208826468 |
| C | 2.370594585872 | -1.161901906850 | 0.608408033756 |
| C | 2.551572900101 | -3.053406542944 | -0.921827487179 |
| C | 3.793194338892 | -3.368458565480 | -0.346476707894 |
| C | 4.315515234352 | -2.586632773114 | 0.695247599346 |
| C | 3.609677269274 | -1.479576712044 | 1.179740779816 |
| H | -5.455584141790 | 1.988442785556 | -1.999050627797 |
| H | -4.309910509456 | 0.008278802428 | -2.945853318944 |
| H | -2.105520571071 | -0.723305284345 | -2.075436944232 |
| H | -4.415655203512 | 3.288065981699 | -0.150142953476 |
| H | -2.070356879362 | 3.434812229294 | 1.584175582337 |
| H | 2.357407247711 | 1.029508512452 | 2.516068778035 |
| H | 0.503084361149 | 2.675689992473 | 2.684819462637 |
| H | 0.501096408583 | -1.112315434107 | -1.946251605008 |
| H | -0.306664902104 | -2.143118963990 | -0.767463772690 |
| H | 2.153787888007 | -3.663303138252 | -1.727823227019 |
| H | 4.353531271394 | -4.223780526291 | -0.710512318882 |
| H | 5.276701747150 | -2.843541320174 | 1.129279598838 |
| H | 4.020275760465 | -0.879500991232 | 1.985954283129 |

**Supplementary Table 6 | DFT-optimized gas phase geometry of 7.**

| Atom | X (Å) | Y (Å) | Z (Å) |
|---|---|---|---|
| C | -4.463433722451 | 1.611654029028 | -1.664386043649 |
| C | -3.790702034001 | 0.486640116448 | -2.163076605258 |
| C | -2.548521077631 | 0.105981347855 | -1.642201879526 |
| C | -1.979676097998 | 0.865482796512 | -0.610178282647 |
| C | -2.664872584464 | 2.003277027704 | -0.109100857066 |
| C | -3.899765579209 | 2.376533026953 | -0.631310857412 |
| C | -0.716577433729 | 0.719352085761 | 0.124191937844 |
| C | -0.623954968623 | 1.768633447154 | 1.078758286441 |
| C | -1.854288357225 | 2.655330911749 | 0.995937533527 |
| C | 0.331043365969 | -0.219734961122 | 0.038302428371 |
| C | 1.462689063781 | -0.078862492708 | 0.927202526738 |
| C | 1.529470173495 | 0.954987499332 | 1.851091732529 |
| C | 0.473589365972 | 1.890110064877 | 1.927998522489 |
| C | 0.549422786206 | -1.368142464194 | -0.789093079013 |
| C | 1.811558595909 | -1.955941128370 | -0.430687445108 |
| C | 2.393805587106 | -1.176432951489 | 0.627448373949 |
| C | 2.486553598499 | -3.082236072209 | -0.928941376375 |
| C | 3.727163147498 | -3.428142215763 | -0.379614603545 |
| C | 4.292009724505 | -2.663143515602 | 0.654366291687 |
| C | 3.626543671300 | -1.531924641899 | 1.163280033419 |
| H | -5.425959981307 | 1.892449952142 | -2.080031003780 |
| H | -4.238056922627 | -0.096026126918 | -2.962261416191 |
| H | -2.042848891733 | -0.766367590174 | -2.039751658099 |
| H | -4.424138071671 | 3.247137611742 | -0.247956749064 |
| H | -1.595703706496 | 3.696144914104 | 0.754591509281 |
| H | -2.405815560554 | 2.683399891509 | 1.946430895343 |
| H | 2.383560640370 | 1.052420109123 | 2.514419443040 |
| H | 0.523090363047 | 2.699911152815 | 2.649759768305 |
| H | -0.111933778860 | -1.743341796303 | -1.557439983697 |
| H | 2.054510817892 | -3.677185620334 | -1.728154116732 |
| H | 4.259182841771 | -4.296185145194 | -0.755602373511 |
| H | 5.254777891001 | -2.947674347760 | 1.066994632341 |
| H | 4.077883319184 | -0.951949093358 | 1.962887011741 |

**Supplementary Table 7 | DFT-optimized gas phase geometry of 8.**

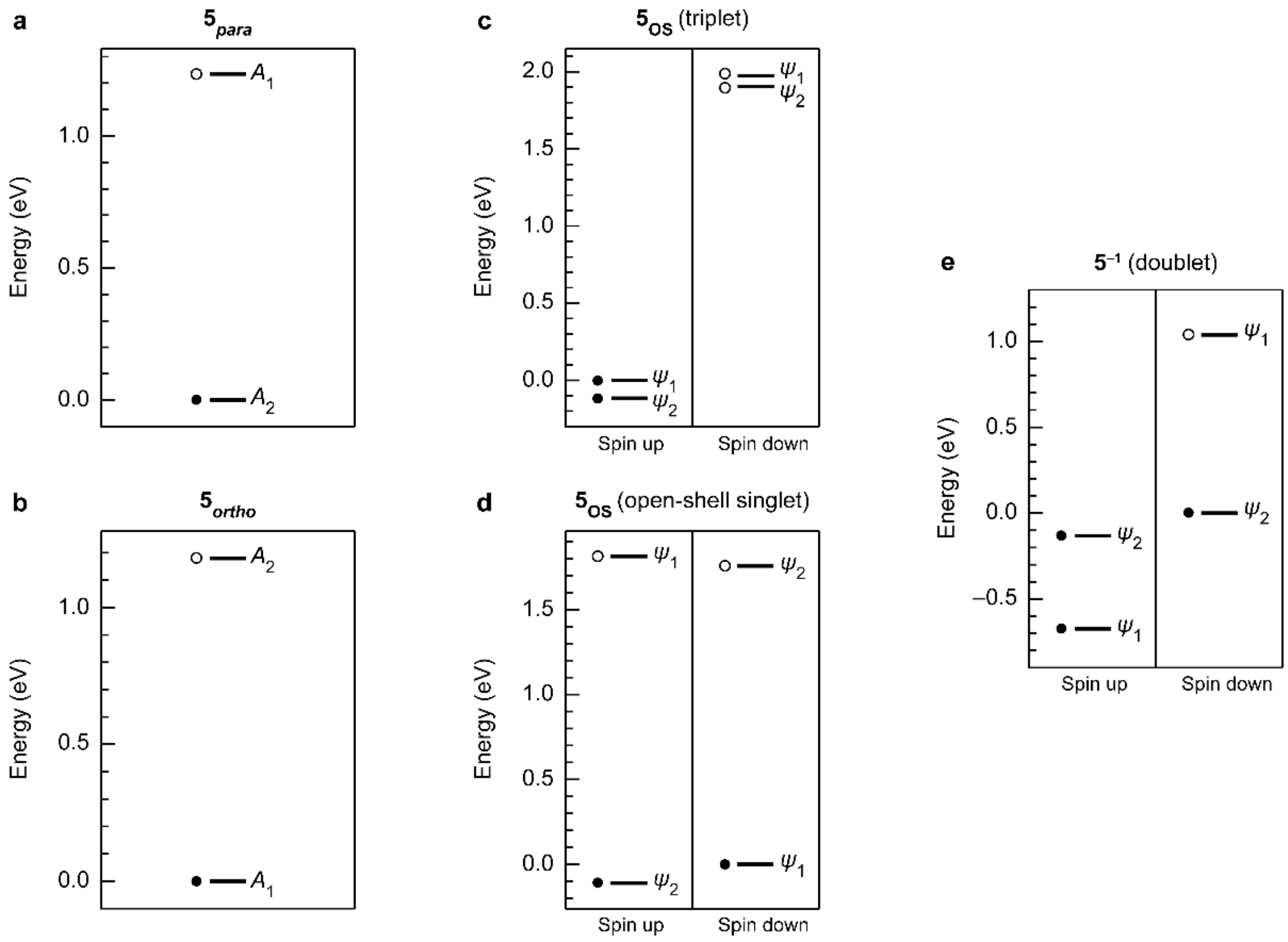

**Supplementary Fig. 5 | Density functional theory calculations on indeno[1,2–*a*]fluorene. a,b,** Frontier orbital spectrum of $\mathbf{5}_{para}$ (**a**) and $\mathbf{5}_{ortho}$ (**b**). For $\mathbf{5}_{ortho}$, $A_1$ and $A_2$ correspond to the HOMO and LUMO, respectively, whereas for $\mathbf{5}_{para}$, $A_2$ is the HOMO and $A_1$ is the LUMO. Wave functions of $A_1$ and $A_2$ are shown in Fig. 3a of the main text. **c,d,** Frontier orbital spectrum of $\mathbf{5}_{OS}$ in the triplet (**c**) and the open-shell singlet (**d**) configurations. $\psi_1$ and $\psi_2$ denote the SOMOs of the open-shell states. Wave functions of $\psi_1$ and $\psi_2$ are shown in Fig. 2a of the main text. For both the triplet and open-shell singlet configurations, $\psi_1$ is the highest-energy occupied orbital and $\psi_2$ is the lowest-energy unoccupied orbital. **e,** Frontier orbital spectrum of $\mathbf{5}^{-1}$. In the anionic state, $\psi_2$ becomes doubly occupied and $\psi_1$ is the SOMO. Filled and empty circles denote occupied and empty orbitals, respectively. For each panel, zero of the energy axis has been aligned to the respective highest-energy occupied orbital.

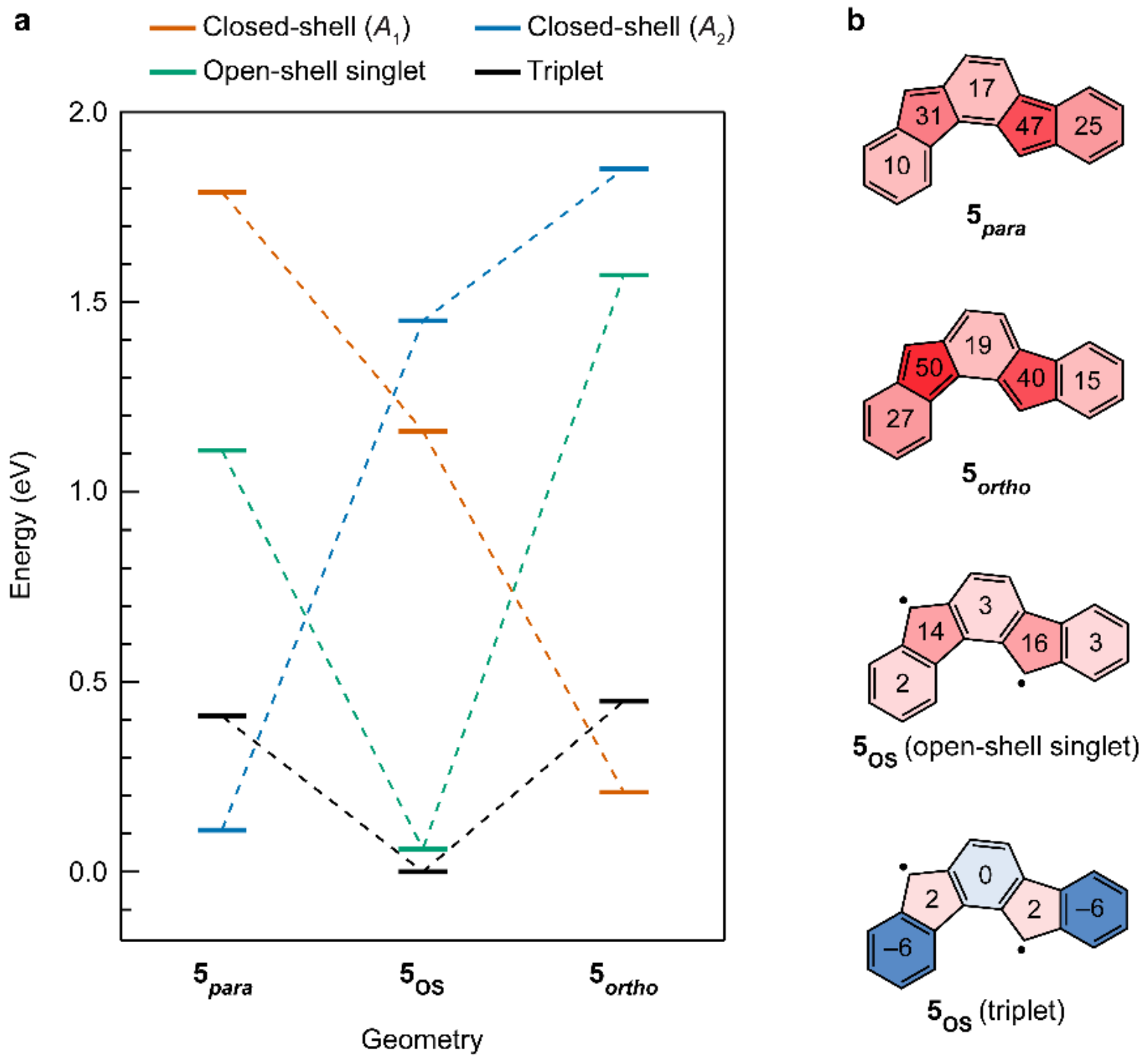

**Supplementary Fig. 6 | QD-NEVPT2 and NICS calculations on indeno[1,2-*a*]fluorene. a,** QD-NEVPT2 energies of four electronic configurations of $\mathbf{5}^{0}$, evaluated at three DFT-optimized geometries labeled on the x-axis, namely, $\mathbf{5}_{\mathrm{OS}}$, $\mathbf{5}_{\boldsymbol{ortho}}$, and $\mathbf{5}_{\boldsymbol{para}}$. The four electronic configurations are, closed-shell ($A_1$): closed-shell configuration with $A_1$ mostly occupied and $A_2$ mostly unoccupied, closed-shell ($A_2$): closed-shell configuration with $A_2$ mostly occupied and $A_1$ mostly unoccupied, triplet: open-shell triplet configuration, with $A_1$ and $A_2$ mostly singly occupied, and open-shell singlet: open-shell singlet configuration, with $A_1$ and $A_2$ mostly singly occupied. Zero of the energy axis corresponds to the state with $\mathbf{5}_{\mathrm{OS}}$ geometry and triplet configuration. We note that all states have multireference character, and the labeling reflects the determinant with the dominant contribution to the multireference wavefunction (the weight of the main determinant being around 60–70%, with the rest accounting for contribution from the other states). DFT geometry optimization of **5** in the triplet and open-shell singlet configurations yields very similar geometries, with the calculated bond lengths of both geometries differing by less than 0.01 Å. Single-point multireference calculations find that in the DFT-optimized triplet geometry (that is, $\mathbf{5}_{\mathrm{OS}}$), both the triplet and open-shell singlet configurations have the lowest energies (the latter is 10 meV below the energy of the open-shell singlet configuration evaluated at the DFT-optimized unrestricted singlet geometry). Therefore, we consider $\mathbf{5}_{\mathrm{OS}}$ as the optimal geometry for both the triplet and open-shell singlet configurations. Given the small energy differences between the triplet and singlet electronic configurations of **5**, low-bias spectroscopic signatures such as spin excitations may be experimentally observed. However, the high mobility of **5** on bilayer NaCl precludes the employment of larger tunneling currents that are typically required to detect inelastic processes such as spin excitations. Note that while we cannot distinguish between the triplet and open-shell singlet electronic configurations of **5** in our experiments, future experiments employing electron spin resonance[6] or alternate-charging[7,8] STM may be able to distinguish between the two configurations. From our (single-reference) DFT calculations, $\mathbf{5}_{\boldsymbol{para}}$ and $\mathbf{5}_{\boldsymbol{ortho}}$ are 0.40 and 0.43 eV, respectively, higher than $\mathbf{5}_{\mathrm{OS}}$, while from multireference QD-NEVPT2 calculations, $\mathbf{5}_{\boldsymbol{para}}$ and $\mathbf{5}_{\boldsymbol{ortho}}$ are 0.11 and 0.21 eV, respectively, higher than $\mathbf{5}_{\mathrm{OS}}$. Thus, compared to DFT calculations, the relative energies of $\mathbf{5}_{\boldsymbol{para}}$ and $\mathbf{5}_{\boldsymbol{ortho}}$ are lowered by 0.29 and 0.22 eV, respectively, in the QD-NEVPT2 calculations. **b,** Results of NICS calculations on **5**. NICS(0)$_{\mathrm{iso}}$ values of **5** in different states are calculated at the B3LYP/def2-TZVP level of theory. Negative (positive) NICS(0)$_{\mathrm{iso}}$ values are indicative of aromaticity (antiaromaticity), while values close to zero suggest nonaromaticity. $\mathbf{5}_{\mathrm{OS}}$ in the triplet configuration exhibits local aromaticity at the terminal benzenoid rings, whereas $\mathbf{5}_{\mathrm{OS}}$ in the open-shell singlet configuration, $\mathbf{5}_{\boldsymbol{para}}$ and $\mathbf{5}_{\boldsymbol{ortho}}$ all display antiaromaticity[9–11]. Thereby, switching between $\mathbf{5}_{\mathrm{OS}}$ (triplet) and $\mathbf{5}_{\boldsymbol{para}}$, as in our experiments, entails a substantial change in aromaticity of **5**.

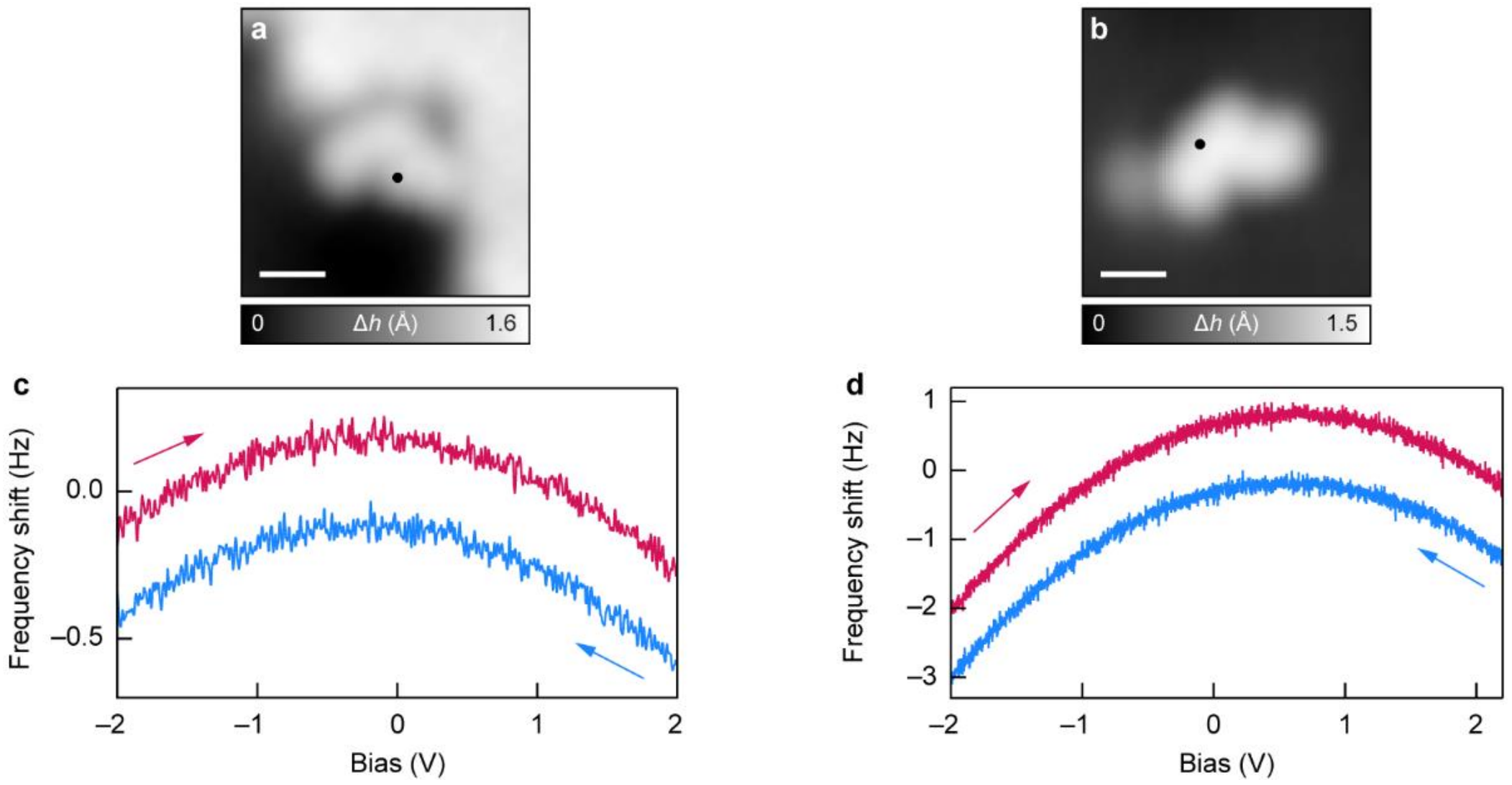

**Supplementary Fig. 7 | Kelvin probe force spectroscopy on indeno[1,2-*a*]fluorene on bilayer NaCl/Au(111). a,b,** In-gap STM images of $\mathbf{5}_{\mathrm{OS}}$ (**a**) and $\mathbf{5}_{para}$ (**b**) on bilayer NaCl/Au(111). The molecule in **a** is shown in Fig. 2, while the molecule in **b** is shown in Fig. 3d in the main text. Scanning parameters: $V$ = 0.2 V, $I$ = 0.3 pA (**a**) and 0.5 pA (**b**). $\Delta h$ denotes the tip height. **c,d,** Constant-height $\Delta f(V)$ spectra acquired on $\mathbf{5}_{\mathrm{OS}}$ (**c**) and $\mathbf{5}_{para}$ (**d**). Open feedback parameters: $V$ = 2 V, $I$ = 0.17 pA (**c**) and $V$ = 2.2 V, $I$ = 0.15 pA (**d**). Colored arrows indicate the bias sweep directions for the corresponding $\Delta f(V)$ curves. Bias sweeps are as follows, forward sweep: 2 V to –2 V (blue) and backward sweep: –2 V to 2 V (red) (**c**); forward sweep: 2.2 V to –2 V (blue) and backward sweep: –2 V to 2.2 V (red) (**d**). The backward $\Delta f(V)$ curve is offset by 0.3 Hz (**c**) and 1 Hz (**d**) relative to the corresponding forward curve for visual clarity. Acquisition positions of the $\Delta f(V)$ spectra are marked by a filled circle in the corresponding in-gap STM images. The acquisition positions shown in **a** and **b** also correspond to the acquisition positions for the $I(V)$ curves in Fig. 2e and Fig. 3c in the main text, respectively. Abrupt steps in $\Delta f(V)$ spectra of atoms[12] and molecules[13] on insulating surfaces have been shown to arise from charge-state transitions. The lack of any steps in the $\Delta f(V)$ spectra in the present case, in conjunction with the absence of NaCl/Au(111) interface-state scattering by **5** (not shown), signifies the neutral charge state of **5** on bilayer NaCl/Au(111) for both $\mathbf{5}_{\mathrm{OS}}$ and $\mathbf{5}_{para}$. Scale bars: 5 Å.

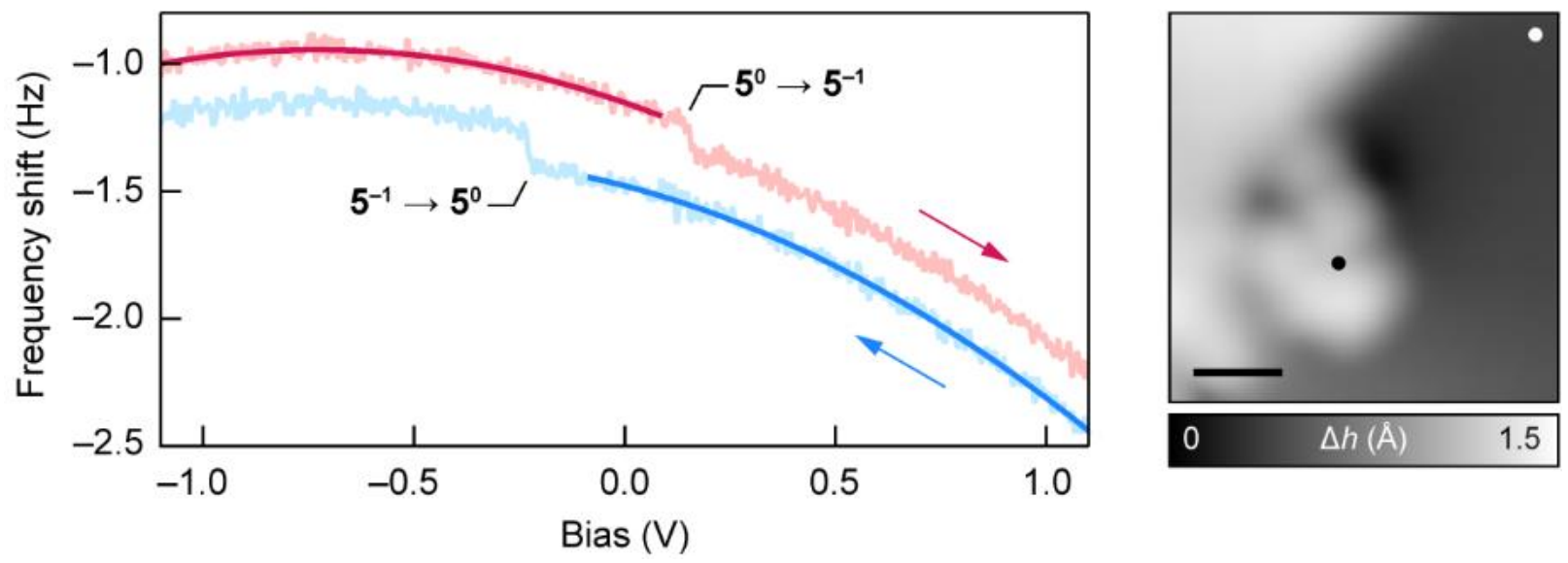

| Δz (Å) | LCPD above $5^{-1}$ (V) | LCPD above $5^{0}$ (V) | LCPD above bilayer NaCl (V) |
|---|---|---|---|
| 0 | –0.529 | –0.727 | –0.866 |
| –0.5 | –0.538 | –0.746 | –0.852 |
| –1.0 | –0.520 | –0.726 | –0.846 |

**Supplementary Fig. 8 | Kelvin probe force spectroscopy on indeno[1,2-*a*]fluorene on bilayer NaCl/Ag(111).** Constant-height $\Delta f(V)$ spectra acquired on **5** on bilayer NaCl/Ag(111). Open feedback parameters: $V$ = 1.1 V, $I$ = 0.2 pA. Forward sweep: 1.1 V to –1.1 V (blue) and backward sweep: –1.1 V to 1.1 V (red). The backward $\Delta f(V)$ curve is offset by 0.2 Hz relative to the forward curve. Also shown are parabolic fits to the $\Delta f(V)$ curves on **5** in the two states. The local contact potential difference (LCPD) is extracted from the abscissas of the vertices of the $\Delta f(V)$ parabolas for three $\Delta z$ values. A larger (more

positive) LCPD indicates a more negatively charged species[14]. As shown in the table, for the entire Δ*z* range, LCPD values above **5** for the state corresponding to *V* > 0.2 V (**5$^{-1}$**) are larger than for the state corresponding to *V* < –0.2 V (**5$^{0}$**). This indicates that the steps in the Δ*f*(*V*) spectra correspond to charge state transitions. The identity of the states as **5$^{-1}$** and **5$^{0}$** is confirmed by the observation of the presence (**5$^{-1}$**) or absence (**5$^{0}$**) of NaCl/metal interface-state scattering by the respective species (Supplementary Fig. 9). LCPD values on bilayer NaCl are shown for reference. The hysteretic behavior of the charging (**5$^{0}$** → **5$^{-1}$**)/discharging (**5$^{-1}$** → **5$^{0}$**) process relates to the reorganization energy[13]. Within the hysteresis loop, **5** is charge bistable and can be imaged both in its neutral and anionic states. Open feedback parameters: *V* = 1.1 V, *I* = 0.2 pA. Acquisition positions of the Δ*f*(*V*) spectra on **5** and on bilayer NaCl (not shown here) are marked by filled circles in the in-gap STM image of **5$^{-1}$** (*V* = 0.2 V, *I* = 1 pA). Note that in its neutral state, this species corresponds to **5$_{OS}$**. Scale bar: 5 Å. On bilayer NaCl/Ag(111) and bilayer NaCl/Cu(111) (Supplementary Fig. 9), **5** exhibits charge bistability. This contrasts with **2**, an isomer of **5**, which was found to adopt a neutral charge state on bilayer NaCl/Cu(111)[3]. This difference results from the different frontier orbital gaps of **2** and **5**, and the relative alignment of the frontier orbitals with respect to the Fermi level of the surface[15].

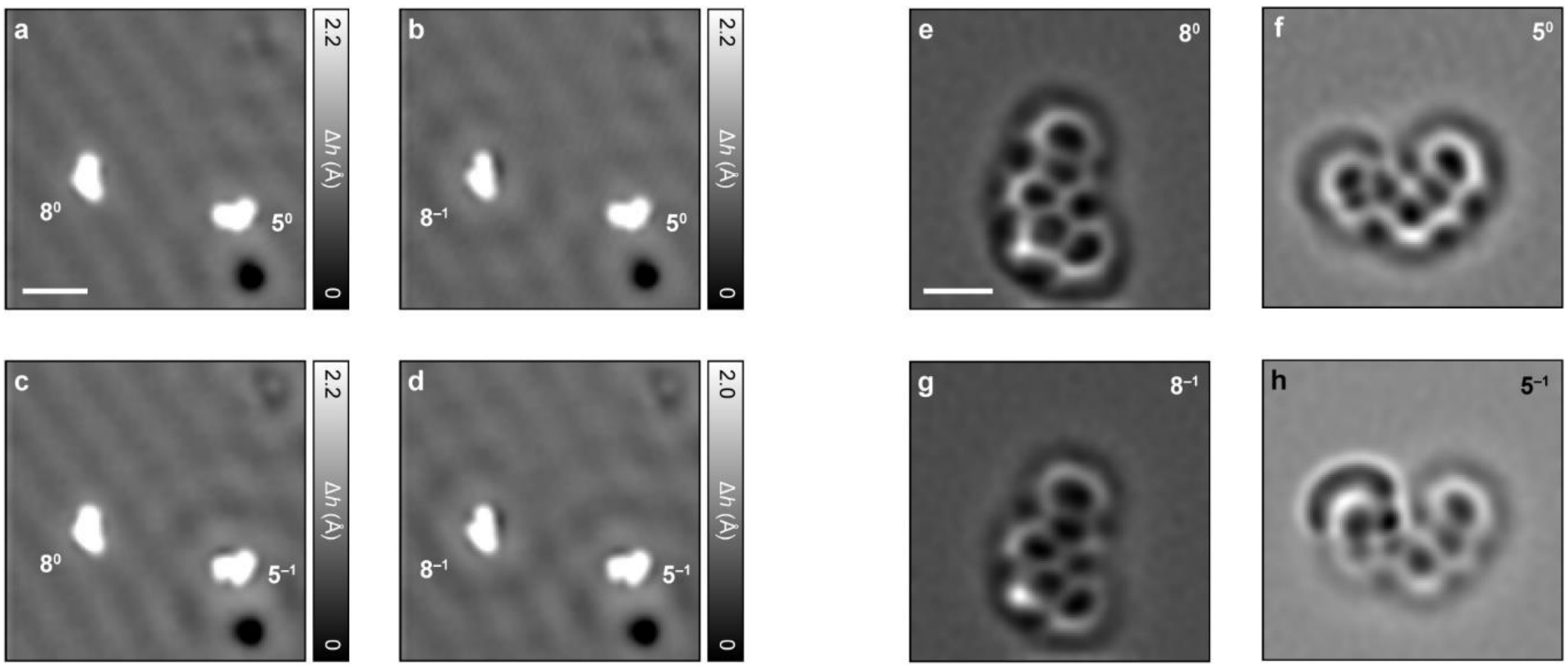

**Supplementary Fig. 9 | Observation of NaCl/Cu(111) interface-state scattering by charged species. a–d,** STM images of **5** and monoradical **8** on bilayer NaCl/Cu(111) in their neutral and negative states: **5$^{0}$** and **8$^{0}$** (**a**), **5$^{0}$** and **8$^{-1}$** (**b**), **5$^{-1}$** and **8$^{0}$** (**c**), and **5$^{-1}$** and **8$^{-1}$** (**d**). Scattering of NaCl/Cu(111) interface state[15,16] is observed by the charged species **5$^{-1}$** and **8$^{-1}$**. Scanning parameters: *V* = 0.1 V, *I* = 0.3 pA. The bias voltage (*V* = 0.1 V) is chosen to be above the onset[16] of NaCl/Cu(111) interface state at approximately –0.2 V. **e–h,** Laplace-filtered AFM images of **5** (**f,h**) and **8** (**e,g**) in their neutral and negative states. STM set point: *V* = 0.2 V, *I* = 0.5 pA on bilayer NaCl, Δ*z* = –0.4 Å. In the absence of any adjacent third layer NaCl island, adsorbates or defects on the surface, **5** moves under the influence of the tip during image acquisition[17]. This causes the apparent bisection of the leftmost ring in the AFM images of both **5$^{0}$** and **5$^{-1}$**. On the defect-free NaCl surface, **5** always showed this movement and in addition, exhibited mobility when increasing the bias voltage to obtain orbital density images. For these reasons, we could not characterize the electronic configuration of **5** on the defect-free NaCl surface. In general, **5** seems to be less stably adsorbed than its isomer **2** (ref. [3]) on NaCl. This may be related to the lower symmetry of **5**, and the symmetry and geometry of **5** not matching well with the NaCl surface[17]. It could also be related to the existence of different metastable adsorption sites and orientations of **5** on NaCl, possibly allowing smaller movement steps for translations and rotations. Scale bars: 20 Å (**a–d**) and 5 Å (**e–h**).

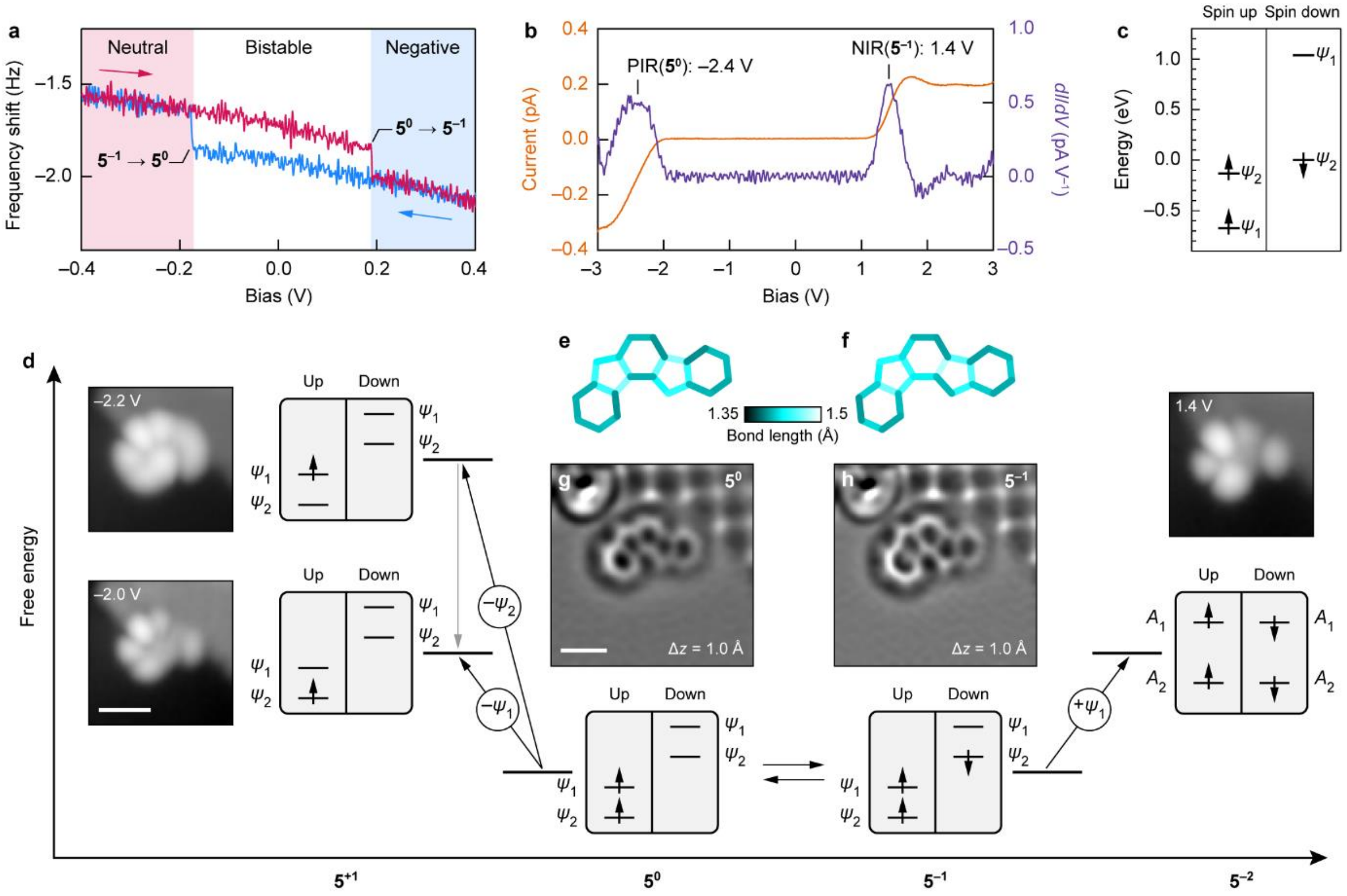

**Supplementary Fig. 10 | Characterization of open-shell indeno[1,2-*a*]fluorene on bilayer NaCl/Ag(111). a,** Constant-height Δ*f*(*V*) spectra acquired on **5** on bilayer NaCl/Ag(111). Open feedback parameters: *V* = 0.4 V, *I* = 0.2 pA. Charge transitions between $\mathbf{5^0}$ and $\mathbf{5^{-1}}$ are indicated. Forward sweep: 0.4 V to –0.4 V (blue) and backward sweep: –0.4 V to 0.4 V (red). **b,** Constant-height *I*(*V*) spectrum acquired on **5**, along with the corresponding *dI/dV*(*V*) spectrum. Open feedback parameters: *V* = 3 V, *I* = 0.15 pA. Given that **5** is in a neutral state for *V* < –0.17 V and in an anionic state for *V* > 0.19 V, the peaks at –2.4 V and 1.4 V correspond to the PIR of $\mathbf{5^0}$ and NIR of $\mathbf{5^{-1}}$, respectively. In the many-body picture shown in (**d**), the NIR corresponds to transitions between $\mathbf{5^{-1}}$ and the dianionic ($\mathbf{5^{-2}}$) state of **5**. **c,** DFT-calculated frontier orbital spectrum of $\mathbf{5^{-1}}$. Zero of the energy axis has been aligned to the highest-energy occupied orbital, namely $\psi_{2\downarrow}$. **d,** Scheme of many-body transitions associated to the measured ionic resonances, along with the STM images of **5** at biases where the corresponding transitions become accessible. STM images at –2 V and –2.2 V show the orbital density of $\psi_1$, and superposition of $\psi_1$ and $\psi_2$, respectively; and the STM image at 1.4 V shows orbital density of $\psi_1$. Scanning parameters: *I* = 0.15 pA (*V* = –2 V and –2.2 V) and 0.2 pA (*V* = 1.4 V). **e,f,** DFT-calculated bond lengths of $\mathbf{5_{OS}}$ (**e**) and $\mathbf{5^{-1}}$ (**f**). **g,h,** Laplace-filtered AFM images of $\mathbf{5^0}$ (**g**) and $\mathbf{5^{-1}}$ (**h**). STM set point: *V* = 0.2 V, *I* = 0.5 pA on third layer NaCl island. Note that in the neutral state as shown in **g**, this species corresponds to $\mathbf{5_{OS}}$. Scale bars: 10 Å (**d**) and 5 Å (**g,h**).

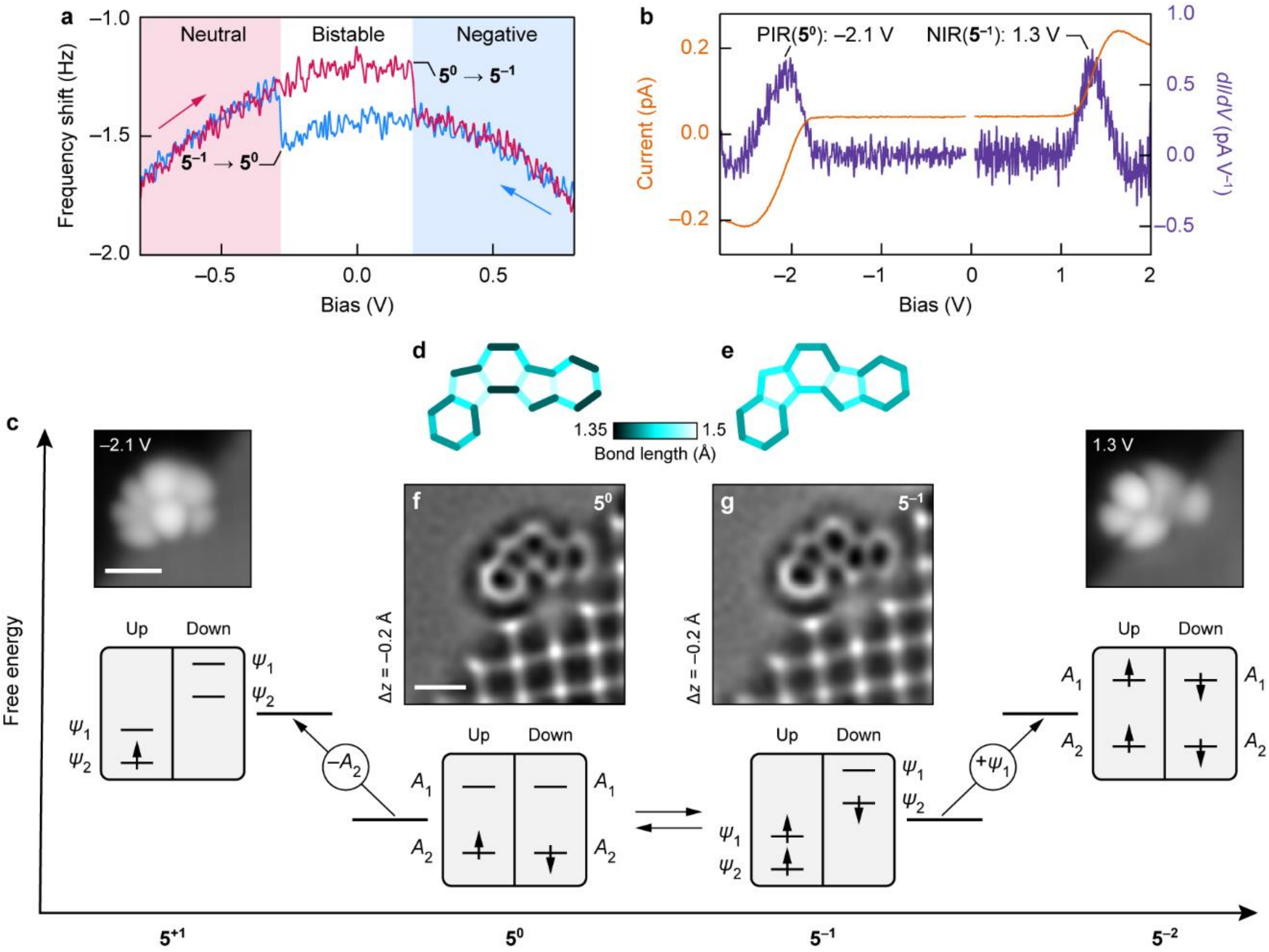

**Supplementary Fig. 11 | Characterization of closed-shell indeno[1,2-*a*]fluorene on bilayer NaCl/Cu(111). a,** Constant-height Δ*f*(*V*) spectra acquired on **5** on bilayer NaCl/Cu(111). Open feedback parameters: *V* = 0.8 V, *I* = 0.2 pA. Charge transitions between $\mathbf{5^0}$ and $\mathbf{5^{-1}}$ are indicated. Forward sweep: 0.8 V to –0.8 V (blue) and backward sweep: –0.8 V to 0.8 V (red). **b,** Constant-height *I*(*V*) spectra acquired on **5**, along with the corresponding *dI/dV*(*V*) spectra. Open feedback parameters: *V* = –2.8 V, *I* = 0.2 pA (negative bias side) and *V* = 2 V, *I* = 0.2 pA (positive bias side). **c,** Scheme of many-body transitions associated to the measured ionic resonances, along with the STM images of **5** at biases where the corresponding transitions become accessible. STM images at –2.1 V and 1.3 V show orbital densities of $A_2$ and $\psi_1$, respectively. Scanning parameters: *I* = 0.25 pA (*V* = –2.1 V) and 0.3 pA (*V* = 1.3 V). **d,e,** DFT-calculated bond lengths of $\mathbf{5}_{\mathbf{\mathit{para}}}$ (**d**) and $\mathbf{5^{-1}}$ (**e**). **f,g,** Laplace-filtered AFM images of $\mathbf{5^0}$ (**f**) and $\mathbf{5^{-1}}$ (**g**). STM set point: *V* = 0.2 V, *I* = 0.5 pA on bilayer NaCl. Note that in the neutral state as shown in **f**, this species corresponds to $\mathbf{5}_{\mathbf{\mathit{para}}}$. Scale bars: 10 Å (**c**) and 5 Å (**f,g**).

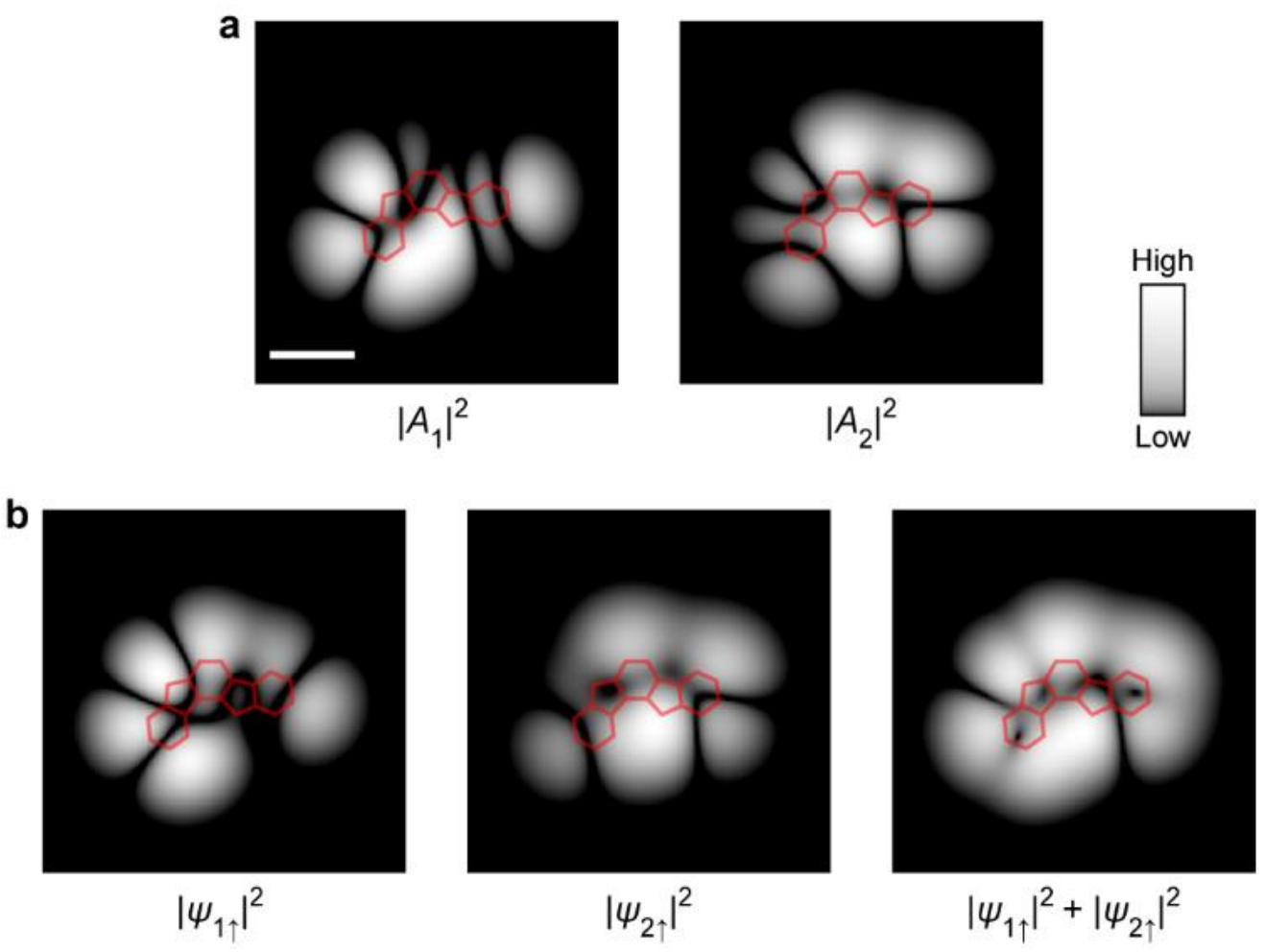

**Supplementary Fig. 12 | Calculated local density of states maps of 5. a,** Constant-height tight-binding local density of states maps of $A_1$ and $A_2$. **b,** Constant-height mean-field Hubbard local density of states maps of $\psi_{1\uparrow}$, $\psi_{2\uparrow}$, and superposition of the densities of $\psi_{1\uparrow}$ and $\psi_{2\uparrow}$. All maps are calculated at a height of 7 Å above the molecular plane and are shown in a logarithmic color scale, which (due to the exponential dependence of the tunneling current on the distance) facilitates comparison with constant-current STM images, shown in Figs. 2 and 3. Scale bar: 5 Å.

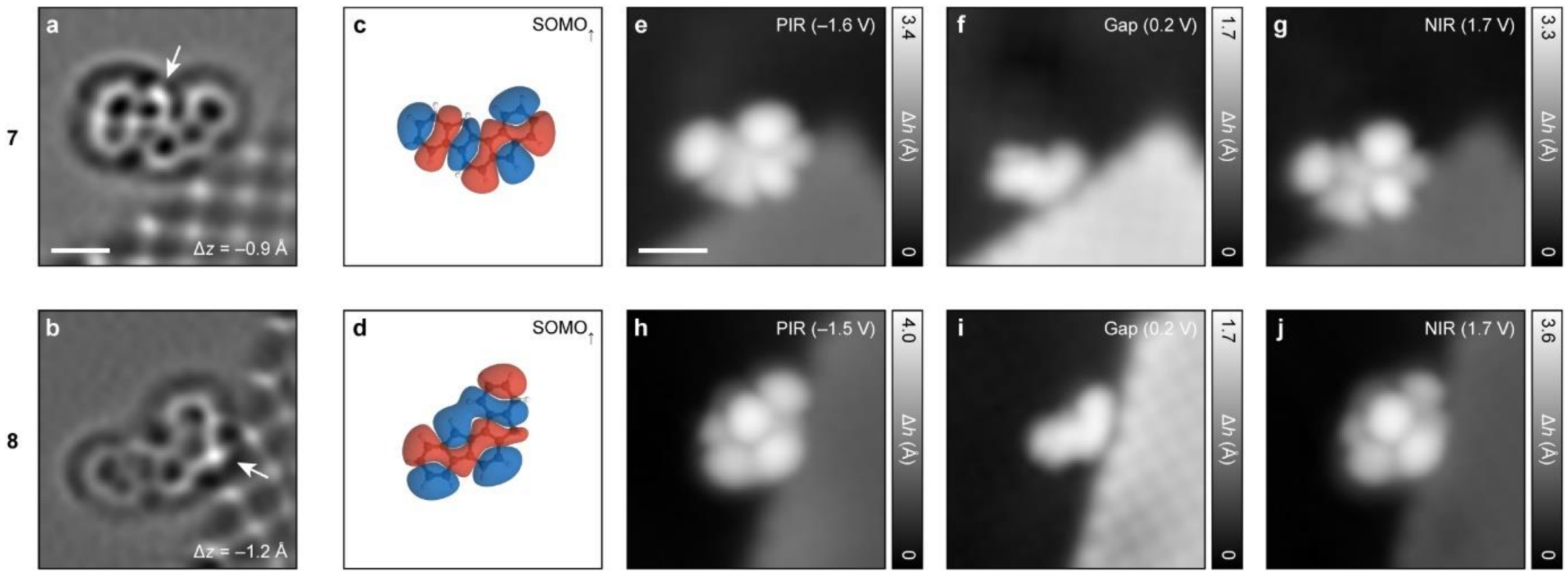

**Supplementary Fig. 13 | Characterization of monoradical species on bilayer NaCl/Au(111). a,b,** Laplace-filtered AFM images of the monoradical species **7** (**a**) and **8** (**b**). Doubly hydrogenated pentagon apexes are indicated with arrows. In these measurements, the monoradical species exhibit movement under the influence of the tip during image acquisition. STM set point: $V$ = 0.2 V, $I$ = 0.5 pA on bilayer NaCl. **c,d,** DFT-calculated wave functions of the SOMO of **7** (**c**) and **8** (**d**) (isovalue: 0.002 $e^-$ $Å^{-3}$). Only the spin up (occupied) level of the SOMOs is shown. DFT calculations predict a doublet ground state of **7** and **8** in the neutral charge state. **e–g,** STM images of **7** acquired at the PIR (**e**), in gap (**f**) and at the NIR (**g**). **h–j,** STM images of **8** acquired at the PIR (**h**), in gap (**i**) and at the NIR (**j**). Scanning parameters: $I$ = 0.15 pA (**e,g**), 0.2 pA (**h,j**) and 0.3 pA (**f,i**). Note that in **f** and **g**, the molecule had translated along the edge of the third layer NaCl island toward the bottom-left of the scan frame. Scale bars: 10 Å (**e–j**) and 5 Å (**a,b**). Gas-phase DFT calculations predict **7** to be roughly 50 meV lower in energy than **8**. Among 17 monoradical species that were generated and analyzed on the metal and NaCl surfaces, eight corresponded to **7** and nine corresponded to **8**, indicating that they are formed with roughly equal probabilities from voltage pulse-induced dehydrogenation of **6**.

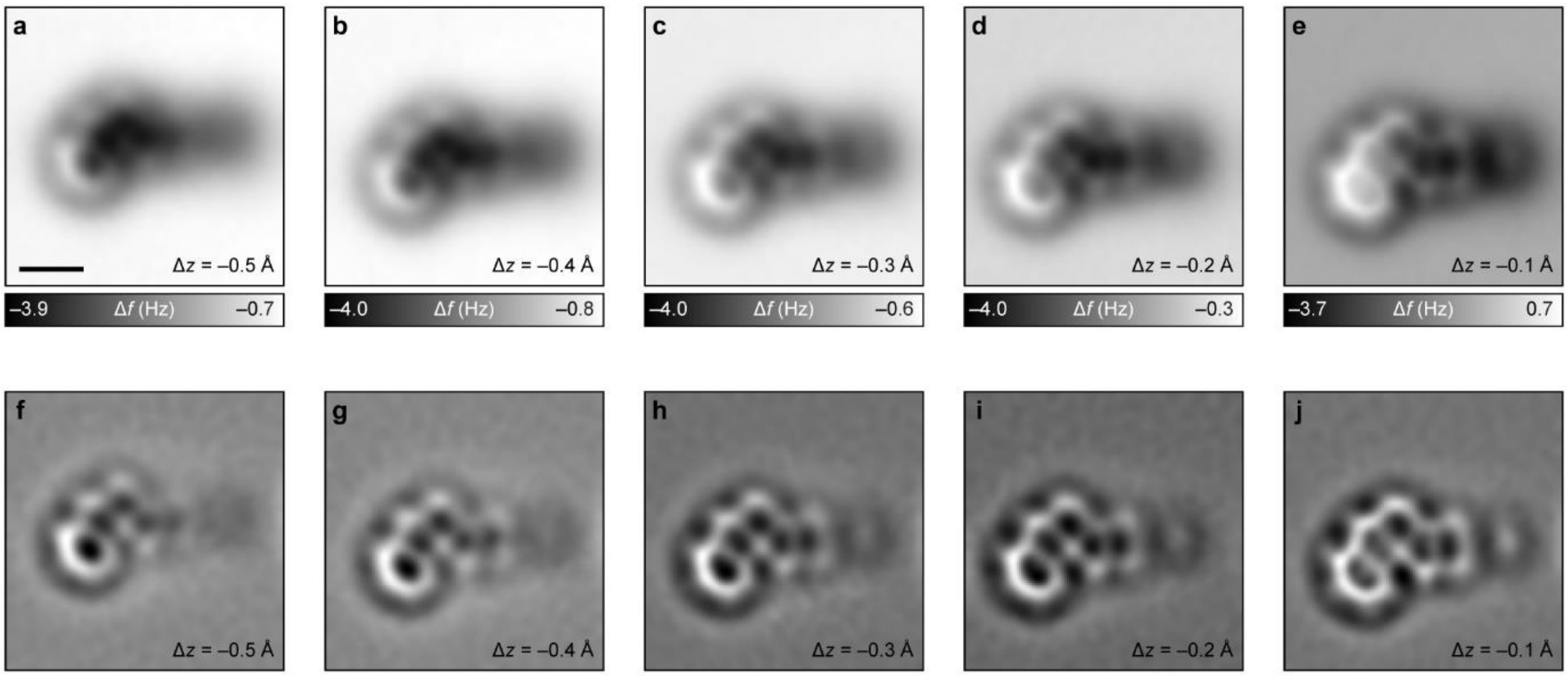

**Supplementary Fig. 14 | Height-dependent AFM imaging of closed-shell indeno[1,2-*a*]fluorene on bilayer NaCl/Cu(111). a–e,** AFM images of $\mathbf{5}_{para}$ at different tip heights. The tip is 0.4 Å closer to **5** in **e** than in **a**. **f–j,** Corresponding Laplace-filtered AFM images. STM set point: $V$ = 0.2 V, $I$ = 0.5 pA on bilayer NaCl. Scale bar: 5 Å.

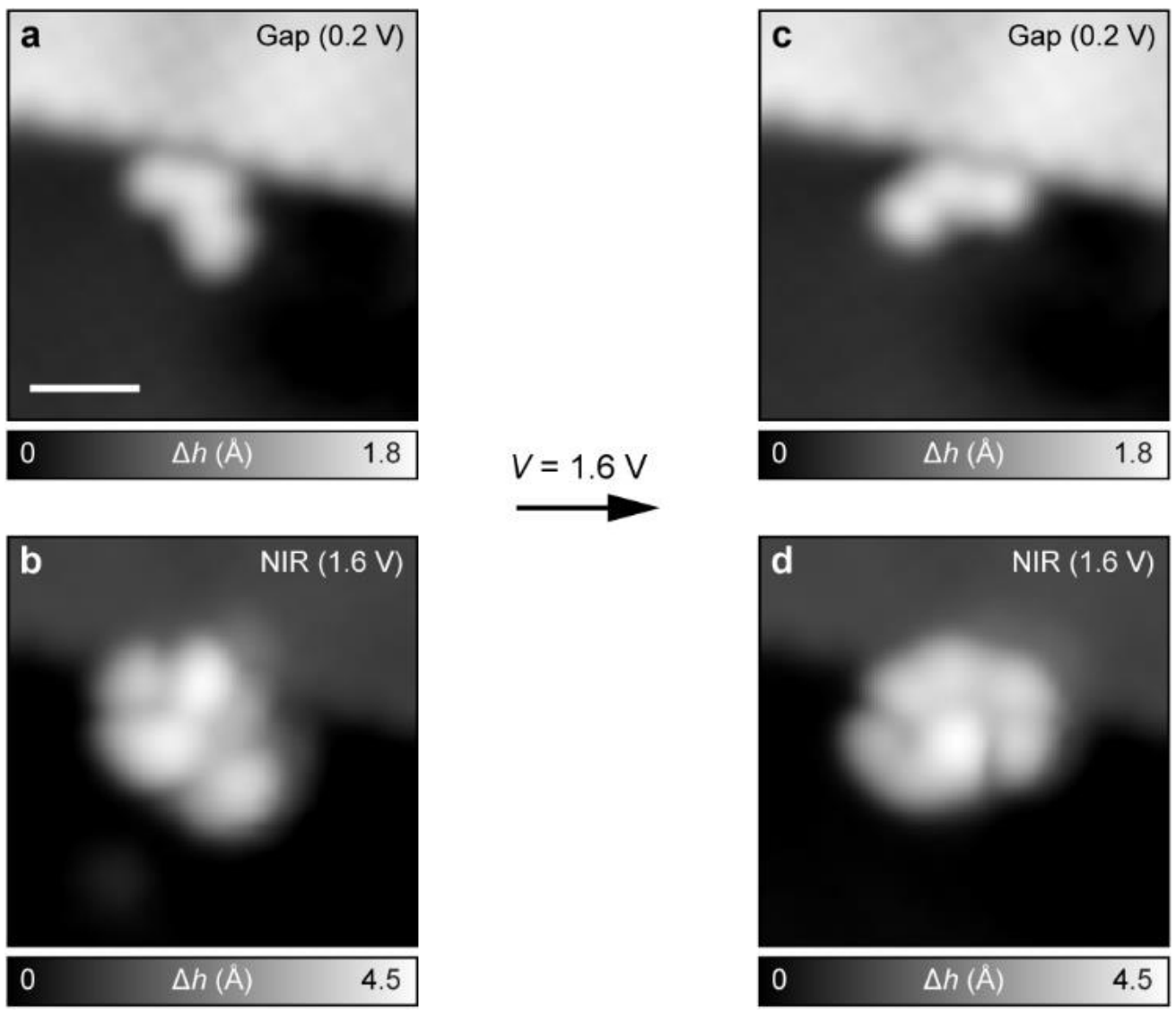

**Supplementary Fig. 15 | Switching between open- and closed-shell states of indeno[1,2-*a*]fluorene on bilayer NaCl/Au(111). a,b,** STM images of $\mathbf{5}_{para}$ acquired in gap (**a**) and at the NIR (**b**). STM image at the NIR shows orbital density of $A_1$, evidencing that the species corresponds to $\mathbf{5}_{para}$. Subsequent scanning at $V$ = 1.6 V led to a change in adsorption site, as observed in the in-gap STM image in **c**. **d,** STM imaging at the NIR now shows superposition of $\psi_1$ and $\psi_2$, evidencing switching to $\mathbf{5}_{\mathrm{OS}}$. Scanning parameters: $I$ = 0.5 pA (**a**), 0.15 pA (**b**), 0.4 pA (**c**) and 0.2 pA (**d**). Scale bar: 10 Å.

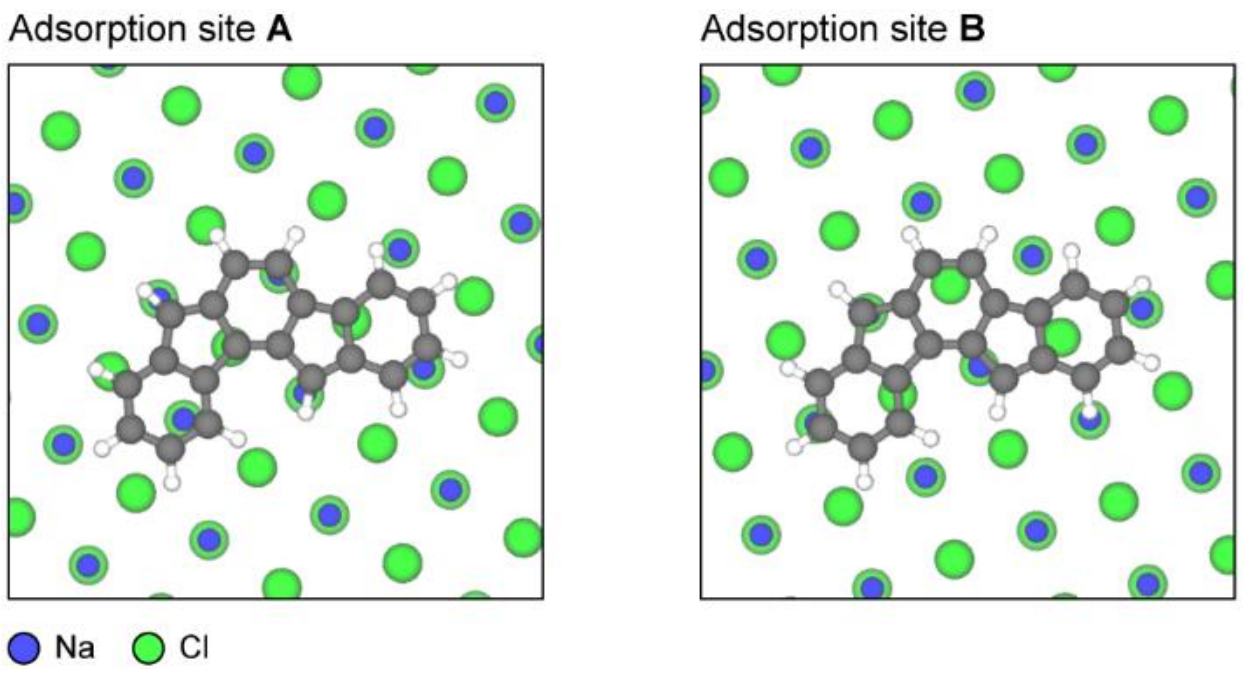

| Adsorption site | Energy, $5_{OS}$ (eV) | Energy, $5_{para}$ (eV) | Energy, $5_{ortho}$ (eV) |
|---|---|---|---|
| **A** | 0 | 0.24 | 0.37 |
| **B** | 0.03 | 0.38 | 0.32 |

**Supplementary Fig. 16 | On-surface DFT calculations.** DFT-optimized adsorption sites of **5** on NaCl (labeled **A** and **B**). The relative energies of $\mathbf{5}_{OS}$, $\mathbf{5}_{para}$ and $\mathbf{5}_{ortho}$ states in sites **A** and **B** are tabulated. The two adsorption sites on bilayer NaCl and the relative stability of the corresponding states in the table were also confirmed with planewave DFT using periodic boundary conditions with one molecule adsorbed on a 5×5 surface slab.

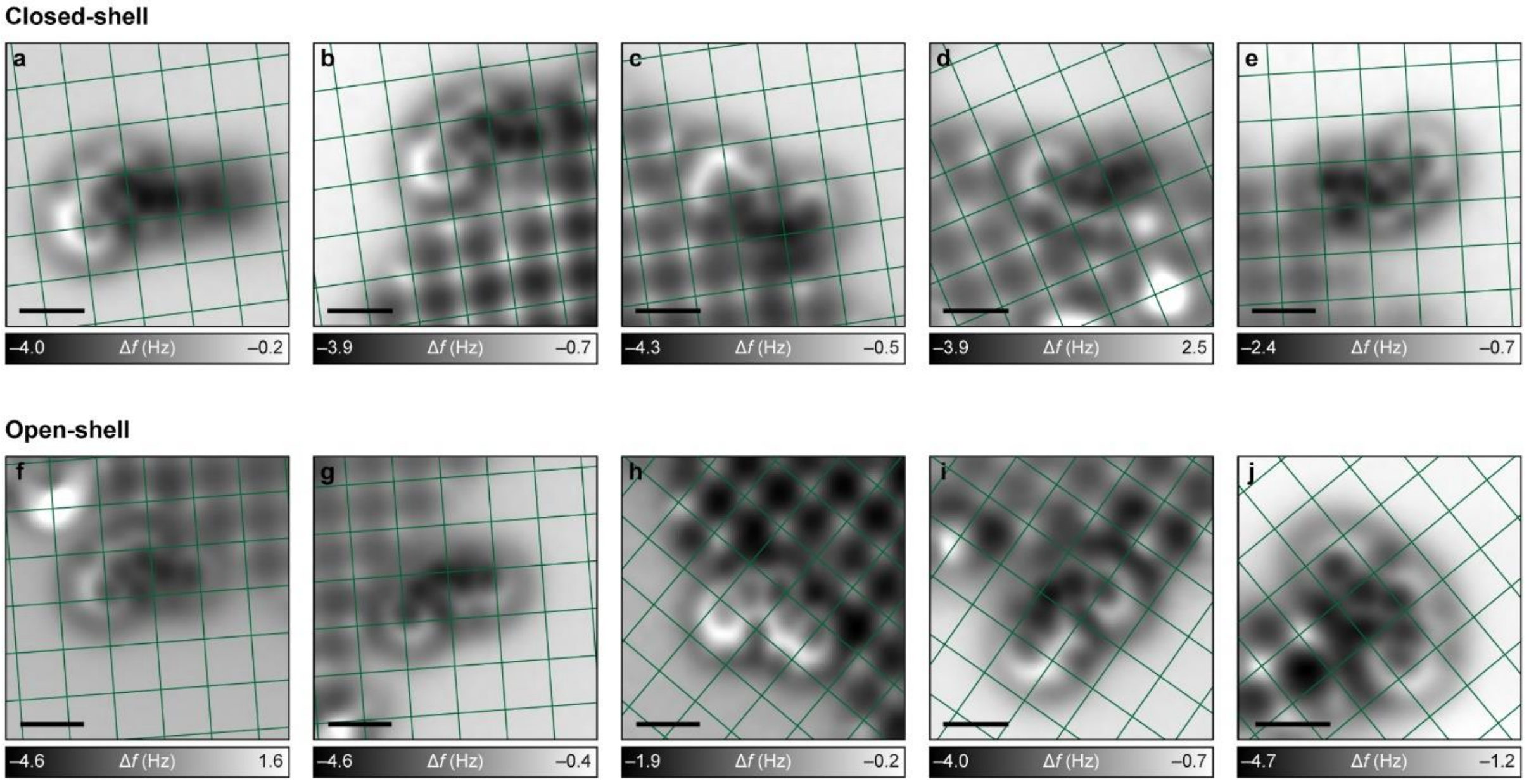

**Supplementary Fig. 17 | Experimental adsorption site determination of 5 using AFM data. a–e,** AFM images of five $\mathbf{5}_{para}$ species. **f–j,** AFM images of five $\mathbf{5}_{OS}$ species. The overlaid lattices (in green) visualize the NaCl lattice. Crossing points correspond to $Na^+$ ($Cl^-$) sites of the second (third) NaCl layer. **5** is adsorbed on bilayer NaCl/Cu(111) in **a**–**d**, **i** and **j**, on bilayer NaCl/Au(111) in **e** and **h**, and on bilayer NaCl/Ag(111) in **f** and **g**. Apart from **a**, where **5** is adsorbed next to a defect on the NaCl surface, **5** is adsorbed next to a third layer NaCl island in all cases. Scale bars: 5 Å.

### Supplementary Note 1. Stabilization of and switching between open- and closed-shell states of 5.

The following experimental observations are made with regards to switching between $\mathbf{5}_{\mathrm{OS}}$ and $\mathbf{5}_{para}$:

1. Switching only takes place when **5** is moved on NaCl, thereby changing its adsorption site. However, in some cases, there is no switching despite movement of **5**.
2. On bilayer NaCl/Ag(111) and Cu(111), where we can reversibly switch the charge state of **5** multiple times between $\mathbf{5}^{0}$ and $\mathbf{5}^{-1}$ (with the $\mathbf{5}^{-1}$ charge state being equivalent, independently of coming from the $\mathbf{5}_{\mathrm{OS}}$ or $\mathbf{5}_{para}$ neutral state), we invariably observe the same neutral ground state ($\mathbf{5}_{\mathrm{OS}}$ or $\mathbf{5}_{para}$) when returning to $\mathbf{5}^{0}$, provided there was no movement of the molecule.

These observations show that by changing the adsorption site of **5** on NaCl, its ground state can be changed, but without changing the adsorption site the ground state of **5** does not change, even if returning from the anionic ($\mathbf{5}^{-1}$) state that is equivalent for both the (neutral) ground states ($\mathbf{5}_{\mathrm{OS}}$ or $\mathbf{5}_{para}$). This leads us to the assumption that stabilization of the open- or closed-shell state of **5** depends on its adsorption site on the NaCl surface. To capture this theoretically, we performed DFT calculations of **5** on a defect-free NaCl(100) surface. Supplementary Fig. 16 shows the DFT-optimized adsorption sites of **5** on NaCl, labeled as **A** and **B**. Sites **A** and **B** are stable and metastable, respectively, for $\mathbf{5}_{\mathrm{OS}}$ and $\mathbf{5}_{para}$. The relative energies of the three states of **5** in site **A** are: $\mathbf{5}_{\mathrm{OS}}$ (0 eV) < $\mathbf{5}_{para}$ (0.24 eV) < $\mathbf{5}_{ortho}$ (0.37 eV); while the energies of the three states in site **B**, relative to the $\mathbf{5}_{\mathrm{OS}}$ state in site **A**, are: $\mathbf{5}_{\mathrm{OS}}$ (0.03 eV) < $\mathbf{5}_{ortho}$ (0.32 eV) < $\mathbf{5}_{para}$ (0.38 eV).

Supplementary Fig. 17 shows the experimentally-determined adsorption sites of five $\mathbf{5}_{para}$ species (Supplementary Fig. 17a–e) and five $\mathbf{5}_{\mathrm{OS}}$ species (Supplementary Fig. 17f–j). For $\mathbf{5}_{para}$, all adsorption sites correspond to site **A**, except the one in Supplementary Fig. 17c that does not correspond to sites **A** or **B** (out of total eight $\mathbf{5}_{para}$ species analyzed, seven were found to adsorb in site **A**). For $\mathbf{5}_{\mathrm{OS}}$, while the adsorption site in Supplementary Fig. 17f corresponds to site **A**, the rest of the adsorption sites do not correspond to sites **A** or **B** (out of total seven $\mathbf{5}_{\mathrm{OS}}$ species analyzed, only one was found to adsorb in site **A**, with the rest not adsorbing in sites **A** or **B**). This poses a conundrum: in site **A**, DFT predicts $\mathbf{5}_{\mathrm{OS}}$ to be the ground state, with $\mathbf{5}_{para}$ 0.24 eV higher in energy; while experimentally, nearly all species found in site **A** correspond to $\mathbf{5}_{para}$, with only one $\mathbf{5}_{\mathrm{OS}}$ species found in site **A**. The resolution to this conundrum may come from more accurate multireference calculations, which, compared to single-reference DFT calculations in gas phase, substantially lower the energies of the closed-shell states relative to the open-shell state (see Supplementary Fig. 6). While multireference calculations are unfeasible on surfaces, we may assume that as in gas phase, on-surface DFT calculations will overestimate the energies of the closed-shell states relative to the open-shell state. Qualitatively, we propose that the DFT-calculated energies of $\mathbf{5}_{para}$ and $\mathbf{5}_{ortho}$ on NaCl are lowered to the extent that $\mathbf{5}_{para}$ would become the ground state in site **A**, that is, by more than 0.24 eV. Note that in site **A**, $\mathbf{5}_{para}$ is found 0.24 eV above $\mathbf{5}_{\mathrm{OS}}$, but in site **B**, $\mathbf{5}_{para}$ is found 0.35 eV (and $\mathbf{5}_{ortho}$ 0.29 eV) above $\mathbf{5}_{\mathrm{OS}}$. Therefore, in site **B**, $\mathbf{5}_{\mathrm{OS}}$ would remain as the ground state if the energies of the closed-shell states are not lowered by more than 0.29 eV. Such magnitude of corrections to single-reference DFT energies seem reasonable, given that the gas-phase energies of the closed-shell states are lowered by 0.22 eV ($\mathbf{5}_{ortho}$) and 0.29 eV ($\mathbf{5}_{para}$) relative to the open-shell state, when using a multireference approach (Supplementary Fig. 6).

These arguments would account for our experimental observation that nearly all $\mathbf{5}_{para}$ species are found in site **A**, while in site **B**, $\mathbf{5}_{\mathrm{OS}}$ would remain as the ground state. Also note that up until now, our discussion from a theoretical standpoint has been focused on **5** on a defect-free NaCl surface. Experimentally, **5** is stably adsorbed only when it is adjacent to a third layer NaCl island, an adsorbate, or a defect on the NaCl surface, each of which will influence the adsorption energetics of **5**. Therefore, deviations from adsorption site **A** for $\mathbf{5}_{para}$ species (as in Supplementary Fig. 17c) can likely be accounted for by the influence of adjacent third layer NaCl islands, adsorbates, or defects on the surface. We find a larger variation in the experimentally-determined adsorption sites of $\mathbf{5}_{\mathrm{OS}}$ (Supplementary Fig. 17f–j) compared to $\mathbf{5}_{para}$, which points toward a shallower adsorption energy landscape of $\mathbf{5}_{\mathrm{OS}}$ on NaCl.

The important finding and observation of on-surface DFT calculations is that the energy differences between $\mathbf{5}_{\mathrm{OS}}$ and $\mathbf{5}_{para}$ change substantially at different adsorption sites. Likely, these changes are even more pronounced for adsorption near third layer NaCl islands, adsorbates, or defects, compared to the defect-free NaCl surface.